\documentclass[aps,onecolumn,superscriptaddress,citeautoscript,notitlepage,10pt,a4paper]{revtex4-1}

\usepackage{pdfpages} 

\usepackage{amsmath, amssymb, bm}
\usepackage{graphicx, epsfig}
\usepackage{textcomp} 
\usepackage{geometry, layouts}
\usepackage[labelfont=bf, figurename=Supplementary~Figure, justification=raggedright, singlelinecheck=false]{caption}
\usepackage{titlesec} 

\makeatletter
\AtBeginDocument{\let\LS@rot\@undefined}
\makeatother

\def\maintextfilename{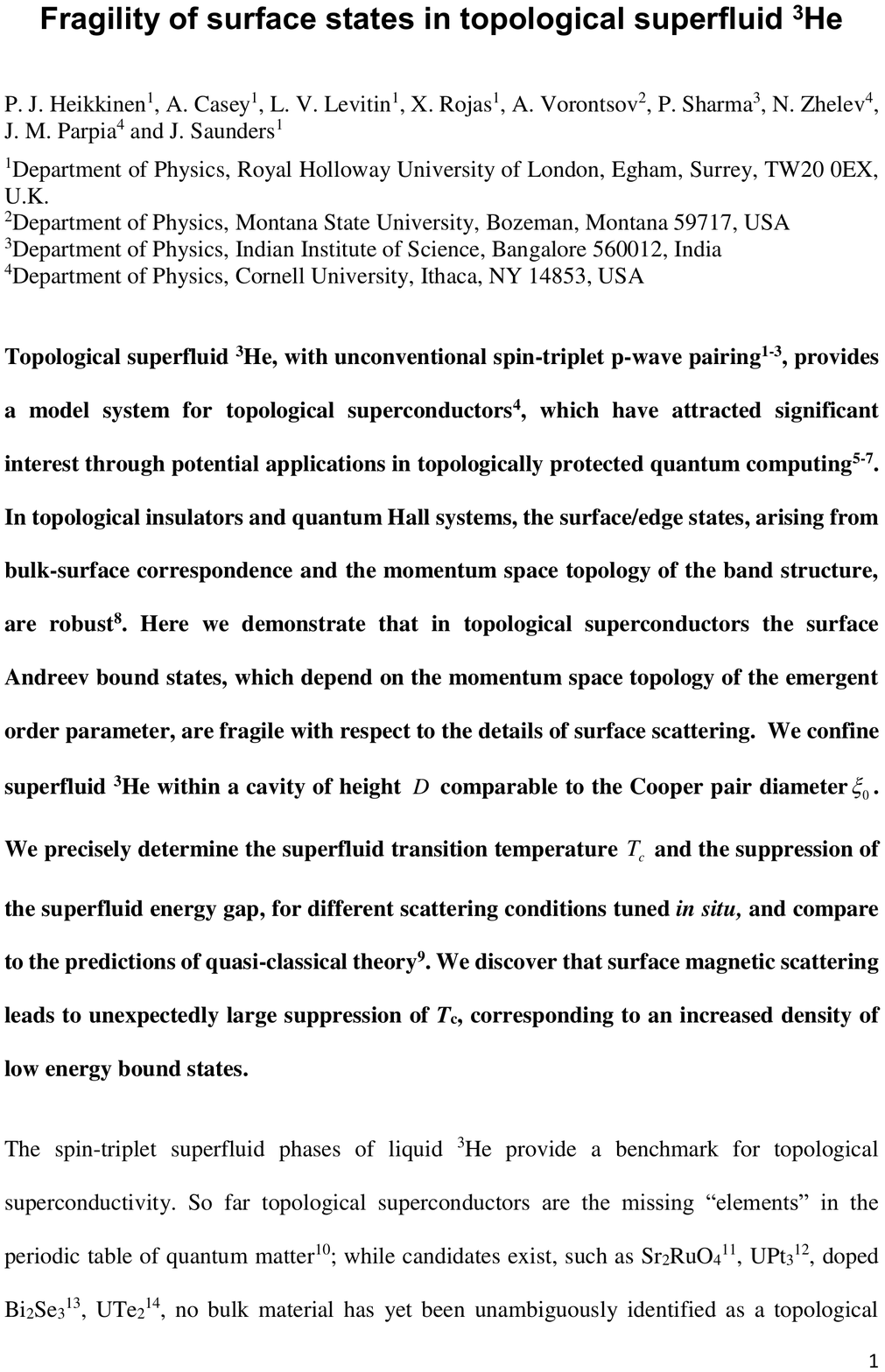}

\titleformat*{\section}{\bfseries}
\titlelabel{SUPPLEMENTARY NOTE \thetitle:\quad}

\begin{document}

\title{Supplementary Information: Fragility of surface states in topological superfluid $^3$He}

\author{P.~J.~Heikkinen}
\email[]{petri.heikkinen@rhul.ac.uk}
\author{A.~Casey}
\author{L.~V.~Levitin}
\author{X.~Rojas}
\affiliation{Department of Physics, Royal Holloway, University of London, Egham, Surrey, TW20 0EX, UK}

\author{A.~Vorontsov}
\affiliation{Department of Physics, Montana State University, Bozeman, Montana 59717, USA}

\author{P.~Sharma}
\affiliation{Department of Physics, Indian Institute of Science, Bangalore 560 012, India}

\author{N.~Zhelev}
\author{J.~M.~Parpia}
\affiliation{Department of Physics, Cornell University, Ithaca, NY 14853, USA}

\author{J.~Saunders}
\affiliation{Department of Physics, Royal Holloway, University of London, Egham, Surrey, TW20 0EX, UK}




# Fragility of surface states in topological superfluid ³He

P. J. Heikkinen[1], A. Casey[1], L. V. Levitin[1], X. Rojas[1], A. Vorontsov[2], P. Sharma[3], N. Zhelev[4], J. M. Parpia[4] and J. Saunders[1]

[1]Department of Physics, Royal Holloway University of London, Egham, Surrey, TW20 0EX, U.K.
[2]Department of Physics, Montana State University, Bozeman, Montana 59717, USA
[3]Department of Physics, Indian Institute of Science, Bangalore 560012, India
[4]Department of Physics, Cornell University, Ithaca, NY 14853, USA

**Topological superfluid ³He, with unconventional spin-triplet p-wave pairing[1-3], provides a model system for topological superconductors[4], which have attracted significant interest through potential applications in topologically protected quantum computing[5-7]. In topological insulators and quantum Hall systems, the surface/edge states, arising from bulk-surface correspondence and the momentum space topology of the band structure, are robust[8]. Here we demonstrate that in topological superconductors the surface Andreev bound states, which depend on the momentum space topology of the emergent order parameter, are fragile with respect to the details of surface scattering. We confine superfluid ³He within a cavity of height $D$ comparable to the Cooper pair diameter $\xi_0$. We precisely determine the superfluid transition temperature $T_c$ and the suppression of the superfluid energy gap, for different scattering conditions tuned *in situ,* and compare to the predictions of quasi-classical theory[9]. We discover that surface magnetic scattering leads to unexpectedly large suppression of $T_c$, corresponding to an increased density of low energy bound states.**

The spin-triplet superfluid phases of liquid ³He provide a benchmark for topological superconductivity. So far topological superconductors are the missing "elements" in the periodic table of quantum matter[10]; while candidates exist, such as Sr₂RuO₄[11], UPt₃[12], doped Bi₂Se₃[13], UTe₂[14], no bulk material has yet been unambiguously identified as a topological

1

crystalline superconductor[15]. Nevertheless devices fabricated from spin-triplet crystalline topological superconductors should eventually contribute to potential applications in topological quantum computation. Current strategies rely on the manipulation of Majorana zero modes (MZMs) in devices which rely on 1-D topological superconductivity[16-19] induced by the proximity effect in topological insulators or semiconductors with strong spin-orbit coupling, or via planar Josephson Junctions[20,21]. However MZMs may be corrupted by non-topological low energy Andreev bound states (ABS)[22], which can be present as a result of details of interfaces and materials properties in such systems[23]. Here we report the fragility of surface ABS in superfluid $^3$He at an ideal non-transparent interface, exploiting the ability to tune *in situ* the scattering of quasiparticles by the surface through adjustment of the isotopic composition of the helium surface boundary layer.

Recently we have shown that it is possible to cool $^3$He confined within precisely engineered nanoscale cavities into the superfluid phases[24], and detect the nuclear magnetic resonance (NMR) response of the small $^3$He sample using an ultra-sensitive spectrometer[25]. Surface scattering dominates the properties under strong confinement. NMR determines the superfluid transition temperature, the pairing state, and the superfluid energy gap[24,26,27], the suppression of which under confinement self-consistently determines, through quasiclassical theory[28], the density of states of the mid-gap surface excitations (ABS).

Superfluid $^3$He consists of spin-triplet Cooper pairs, with one unit ($l = 1$) of orbital angular momentum[1]. The order parameter is a complex 3x3 matrix, encoding the spin state over the spherical Fermi surface. In bulk liquid two phases are found with distinct symmetries[1,2] and momentum-space topologies[3]. The A phase is chiral, breaking time reversal symmetry. Over the Fermi surface pairs form with the same direction of their orbital angular momentum, and in an equal-spin state comprising just $\left|\uparrow\uparrow\right\rangle$ and $\left|\downarrow\downarrow\right\rangle$ pairs. The B-phase is time-reversal invariant, comprising all three components of the spin triplet, with broken spin-orbit relative

2

symmetry. The relative stability of these phases is profoundly altered by confinement.[24] Superfluid $^3$He is an intrinsically impurity-free system, although impurities can be artificially introduced using silica aerogels of different porosities and structure factor[29]. In the work reported here we determine the influence of surface scattering alone on gap suppression, in the absence of impurity scattering.

In our experiment, superfluid $^3$He was confined within a 192 nm high cavity defined in a silicon wafer, Fig. 1a. The effective confinement can be varied at fixed cavity height $D$ by changing the sample pressure and hence the superfluid coherence length $\xi_0 = \hbar v_F / 2\pi k_B T_{c0}$, where $v_F$ is the Fermi velocity and $T_{c0}$ the bulk superfluid transition temperature. Measurements were made at a series of pressures from 0 to 5.5 bar, over which $\xi_0$ decreases from 77 to 40 nm. We determine the shift in the NMR resonance frequency $f$ relative to the Larmor frequency $f_L$, $\Delta f = f - f_L$, which occurs in the superfluid state. The onset of this shift identifies $T_c$ in the cavity. This is determined precisely relative to $T_{c0}$ by also observing frequency shifts in small volumes of bulk liquid incorporated in the cell design, Fig 1a and Supplementary Figure 1. The frequency shifts from the cavity and the bulk markers are of opposite sign, Fig 1b. The superfluid transition within the cavity is sharp, due to the uniformity of cavity height, relative to that achieved in stacked multiple films with a broad distribution of thickness[30]. Suppression of the gap by confinement is inferred from the magnitude of the cavity signal frequency shift.

The relatively strong confinement in the 192 nm cavity stabilizes the A phase at all temperatures and pressures, consistent with the phase diagram determined in previous work[24,30,31]. The orbital angular momentum of the pairs, which defines the orientation of point nodes of the gap in momentum space, lies normal to the cavity surface $\hat{\mathbf{l}} = \pm\hat{\mathbf{z}}$. The order parameter $\boldsymbol{\Delta}(\mathbf{p}) = \Delta_A(z)(\hat{p}_x + i\hat{p}_y)\left[\left|\uparrow\uparrow\right\rangle + \left|\downarrow\downarrow\right\rangle\right]$[31], where $\Delta_A$ is the A-phase gap maximum,

3

which in general has a spatial dependence across the cavity. The static magnetic field $\mathbf{H}_0 = H_0 \hat{\mathbf{z}}$ orients the spins along $\hat{\mathbf{z}}$, via the anisotropic magnetic susceptibility. With this relative orientation of spin and orbital angular momentum, the dipolar energy is maximized, accounting for the negative frequency shift observed from the cavity, Fig. 1 (and Supplementary Note 1).

In quasiclassical theory the surface scattering can be characterized in terms of a single parameter: specularity $S$, Fig. 1d. There is prior compelling evidence that the surface scattering may be tuned *in situ* from diffuse to specular[30,32-35] by coating surfaces with $^4$He, both from hydrodynamic studies of surface slip[32] in normal state $^3$He and from transverse acoustic impedance studies[34,35] of surface excitations in $^3$He-B. These studies suggest close to specular scattering in the presence of a superfluid $^4$He surface film. Quasiclassical theory predicts for the A-phase that $T_c$ and gap suppression, and surface bound states are all eliminated for specular scattering. For diffuse scattering the suppression of $T_c$ scales as $\delta T_c / T_{c0} \sim -(\xi_0 / D)^2$ to leading order. Details of the treatment of surface scattering used in our quasiclassical computations are given in Supplementary Note 3.

We first discuss measurements in which the sample walls and heat exchanger surfaces were plated with sufficient $^4$He to displace the magnetic solid $^3$He surface boundary layer, which arises naturally in pure $^3$He samples[36]. The plating procedure results in a non-magnetic localized solid $^4$He surface boundary layer. In this case the observed $T_c$ suppression is close to that predicted for purely diffuse scattering. The results are best fit with specularity $S = 0.1$, referred to here as "diffuse", Fig. 2a. The increase in $T_c$ suppression with decreasing pressure arises naturally from stronger effective confinement [31], Fig. 2b. Subsequently a thicker $^4$He film was formed on the cavity walls to create a surface superfluid film of $^4$He. In this case we observe an almost complete elimination of $T_c$ suppression, demonstrating close to fully specular scattering, referred to here as "specular", Fig 2a,b.

In general, the measured frequency shift is related to the spatial average of the suppressed gap within the cavity via $\left| f^2 - f_L^2 \right| = \zeta \left\langle \Delta_A^2(z) \right\rangle$, where $\zeta$ is an intrinsic material parameter which is pressure dependent but temperature independent (Supplementary Note 1). In the Ginzburg-Landau regime, sufficiently close to $T_{c0}$, the A-phase bulk gap maximum $\Delta_A$ is given by $\Delta_A^2 = \dfrac{\Delta C_A}{C_n} \left( \pi k_B T_{c0} \right)^2 (1 - T / T_{c0})$, where $\Delta C_A / C_n$ is set to the measured heat capacity jump at $T_{c0}$. This expression thus incorporates strong-coupling corrections to the gap near $T_{c0}$ (Supplementary Note 4). For "specular" boundaries, the measured cavity frequency shift corresponds to the unsuppressed, bulk gap, Fig 2c, and allows determination of the constant $\zeta$ at each pressure. For the "diffuse" boundary, using the determined value of $\zeta$, we can precisely infer the gap-suppression from the measured frequency shift, independent of uncertainties in material parameters and temperature scale, Supplementary Note 1. We find that the observed gap suppression is also best described by $S = 0.1$, establishing the consistency of the experimentally determined gap suppression and $T_c$ suppression within the framework of quasi-classical theory.

We now turn to the results where no [4]He preplating was deployed, leaving a magnetic surface boundary layer of localized [3]He. Rapid exchange with the liquid results in a single hybridized NMR line[30]. The superfluid transition temperature is inferred from analysis of the frequency shift of the hybridized line, which is a weighted average of the internal dipolar frequency shift in the solid [3]He surface boundary layer and that due to superfluidity[30] (Supplementary note 5). It shows an unexpectedly large $T_c$ suppression, Fig. 3a, significantly exceeding that observed with a solid [4]He boundary layer, and inconsistent with diffuse scattering $S \approx 0$. This result can be *phenomenologically* described in terms of an effective specularity $S_{\text{eff}} = -0.4$. This approaches the condition for maximal pair-breaking $S = -1$, corresponding to full retro-reflection, in which case the phase shift φ experienced by the retro-

reflected quasiparticle is $\varphi = \pi$ for all incoming/outgoing trajectories and surface bound states accumulate at zero energy, since $E/\Delta = \pm\cos(\varphi/2)$ [28] (see density of states calculation as a function of specularity in Supplementary Fig. 8). However momentum scattering with a preponderance of retroreflection is inconsistent with measurements of boundary slip in viscous transport in the normal state, which find equivalent specularity ($0 < S < 1$) for both solid $^3$He and solid $^4$He surface boundary layers [32].

We invoke magnetic surface scattering to explain this stronger $T_c$ suppression in the presence of a magnetic solid $^3$He surface boundary layer. Exchange interaction between quasiparticles and isolated magnetic impurities has been theoretically established to induce additional bound states in superconductors through the Yu-Shiba-Rusinov mechanism [37]. Magnetic scattering by localized $^3$He [38] strongly influences the observed superfluid phase diagram of $^3$He in different aerogels [39,40]. Here we extend these ideas to consider exchange scattering by the uniform 2D surface layer. We seek processes which generate an excess of zero-energy states over that found for diffuse momentum scattering (Fig. 3b and Supplementary Note 6).

The structure of the order parameter is such that the phase shift φ experienced by the scattered quasiparticle, and hence the energy of the surface bound states, will be influenced by spin-dependent scattering processes. To account for extra pair breaking we need to include quantum spin dynamics of randomly oriented localized quantum spins, allowing for their spin flips. Exchange coupling between such spins and incident quasiparticles gives rise to interference between the singlet and triplet scattering channels. In our experimental configuration we find this to be the only viable mechanism resulting in enhanced $T_c$ suppression. We find that for a surface with momentum scattering specularity $S$, the suppression of $T_c$ corresponds to an effective specularity between bounds $-S \leq S_{\mathrm{eff}} \leq S$,

depending on strength of exchange coupling. Thus this process is only detectable for non-diffuse surfaces, but can give rise to $T_c$ suppression exceeding that for a diffuse surface, as observed. In order to reach the detected $S_{eff} = -0.4$, the required underlying specularity for momentum scattering from the atomically smooth silicon surface with solid $^3$He surface boundary layer should be $S \geq 0.4$, which is a plausible scenario (Methods).

The specular surfaces we have demonstrated open the investigation of cavities of arbitrarily small height towards $D << \xi_0$ entering the quasi-2D limit, in which thermal and spin analogues of the Quantum Hall effect are predicted[41,42]. Magnetic $^3$He boundaries may stabilize new order parameters under confinement, and influence surface spin currents. Future topological superfluid $^3$He mesoscopic devices should provide a new platform for the study of MZMs at well-defined interfaces. The sculpture of the superfluid by confinement will allow the fabrication of hybrid devices based on different $^3$He "materials", with clean transparent interfaces, Fig. 4. Design of these platforms is supported by quasi-classical theory, which self-consistently describes the spectrum of surface states, gap suppression and $T_c$ suppression.

In conclusion, our results show experimentally the sensitivity of ABS to the details of quasiparticle scattering from the surface. A superfluid $^4$He surface boundary layer results in specular scattering, which eliminates surface ABS in chiral superfluid $^3$He-A. A surface boundary layer of solid $^4$He leads to diffuse scattering, with a finite density of low energy ABS, and superfluidity will be completely suppressed in cavities thinner than 100 nm at zero pressure. We discover that surface magnetic exchange scattering from a solid $^3$He surface boundary layer must further increase the density of low energy ABS, and thus gives rise to a greater suppression of $T_c$ than for diffuse scattering. We predict that this effect is strongest when the momentum scattering is specular. The surface states of $^3$He-A are not topologically protected, unlike its edge states[43] or the Majorana surface states of $^3$He-B[44]. The calculation of

the influence of spin-flip magnetic boundary scattering on the density of states of surface excitations in these two topological superfluids remains an open problem.

In more complex topological superconductor architectures designed to realize, detect and manipulate MZMs for topological quantum computation, it is necessary to eliminate the excess low energy ABS arising from the interface scattering processes such as we have identified here. Meanwhile our result is a crucial step in the quest to produce and identify Majoranas in liquid $^3$He, the as yet only firmly established topological superconductor/superfluid. More generally, the influence of magnetic degrees of freedom in topological materials[45] and spin-active interfaces in hybrid superconducting-ferromagnetic spintronics[46] are both areas of current interest to which the understanding of spin-dependent surface scattering in a spin-triplet superfluid will contribute.

## Methods

**Silicon nanofluidic cavity fabrication.** The experimental cell was fabricated by direct wafer bonding of two silicon wafers. The confinement region and supporting pillars are defined lithographically on one of the wafers using a process similar to that used in a previous generation of cells[47]. The typical surface roughness of the silicon surface is 0.1 nm[48]. This is significantly smoother than the mechanically polished silicon surfaces for which surface specularity has been characterized by normal state studies of slip in viscous transport[32], potentially promoting specularity of surface scattering even in the absence of a superfluid $^4$He film. Deep Reactive Ion Etch (DRIE) is used to create two 300 micron diameter holes either side of the confinement region. One acts as a fill line and the other provides a region of bulk helium on the far side of the slab shaped cavity, Fig. 1a. DRIE is also used to pattern the backside of the wafer to improve the joint between the cell and an external fill line[48]. After all the features are patterned onto the wafers they are cleaned using a combination of a two-step RCA clean at 75 °C followed by immersion in concentrated (49%) HF to remove any oxide or contaminants. The clean patterned wafer is brought into contact with a blank silicon wafer within a wafer aligner, forming a bond between the cavity wafer and the lid. The bond strength is increased and made permanent by an annealing step at 1000 °C for 2 hours. Successful bonding is confirmed using infra-red imaging and scanning acoustic microscopy. The bonded wafer is diced into individual cells using a diamond saw. A 500 nm thick silver film is evaporated onto the outside of the bonded wafers to thermalize the cell to the nuclear stage. In order to minimise the effects of differential thermal contraction between the metallic fill line/ far-end bulk marker plug and the silicon cell, laser-machined silicon washers are attached around both of the DRIE holes with epoxy (Stycast 1266 mixed with silicon powder), Supplementary Fig. 1. The height of the cavity used in this work was determined to be 192 nm by a profilometer scan on the unbonded wafer. The error in the cavity thickness ±2nm was estimated from

the distribution in height measured in this way across all the cavities on the unbonded wafer. The dependence of cavity height on pressure is determined by finite element method simulations to be 2.6 nm/bar, Supplementary Fig. 2.

**NMR measurements**. The cooling of the $^3$He within the cell, the thermometry and the SQUID NMR spectrometer were as used in previous work[24,25], Supplementary Fig. 1. The helium is cooled via the column of $^3$He in the fill line which connects the cell to a sintered silver heat exchanger mounted on a silver plate, connected via a silver rod to the copper nuclear demagnetization stage. A platinum NMR thermometer is mounted on the silver plate. Measurements were made at a $^3$He Larmor frequency of 967 kHz, with the static field of around 30 mT applied along the cavity surface normal ($\hat{\mathbf{z}}$). Field gradients were applied to both separate the bulk marker signals from the cavity signal (along $\hat{\mathbf{z}}$), and to resolve the signals from the two bulk markers (along $\hat{\mathbf{x}}, \hat{\mathbf{y}}$). The free induction decay following small angle tipping pulses, applied at 10s intervals, was averaged typically 30 times, and Fourier transformed. Measurements with tipping pulses of different amplitude enabled a correction to be made for temperature gradients between the $^3$He in the cell and the platinum thermometer (Supplementary Note 7). This correction depended on the surface boundary condition, which influenced the boundary resistance of the silver heat exchanger. The temperature gradient across the cavity is small, and dependent on surface boundary condition; it is determined from the difference between the measured superfluid transition temperature in the two bulk marker volumes. For solid $^4$He and $^3$He surface boundary layer the difference is around 20μK, while for the superfluid $^4$He surface boundary layer it is at most 2μK, Supplementary Fig. 6. This gradient is taken into account in determining the error in superfluid transition temperature.

 **In situ tuning of surface scattering.** *Magnetic scattering.* Pure $^3$He, with $^4$He impurity concentration less than 100 ppm, is used to fill the empty cell and silver heat exchanger (surface area 8 m$^2$). This results in a magnetic surface boundary layer of solid $^3$He[36]. The results suggest that this self-assembled solid $^3$He layer on the extremely smooth silicon surface gives rise to partially specular momentum scattering which combines with spin-flip magnetic scattering. *Diffuse non-magnetic scattering*. In order to displace the naturally occurring magnetic surface boundary layer of $^3$He[36], 32 $\mu$mol/m$^2$ of $^4$He was added to the empty cell and silver heat exchanger at 30K, followed by cooling to below 1K over 30 hours, and a subsequent anneal at 2K for several hours. This coverage is below that necessary to see a superfluid transition in the surface $^4$He layer, in the presence of an overburden of $^3$He at saturated vapour pressure[49]. The sample is cooled to 100 mK before adding $^3$He. Under these conditions the $^3$He surface magnetism seen in pure $^3$He samples is eliminated. The results find that the remaining momentum scattering is close to diffuse, suggesting that the $^4$He pre-plating surface layer has a greater roughness than the self-assembled solid $^3$He boundary layer occurring in pure $^3$He sample. *Specular scattering*. To create specular scattering conditions from the previous $^4$He surface plating conditions, the cell is pumped at 1.5K, leaving a residual solid $^4$He "layer" on the surfaces. Then more $^4$He is added into the cell/heat exchanger. Subsequently the helium pumped out in the previous step is restored. The sample is slowly cooled into the mK range, during which all the $^4$He forms a surface film of solid $^4$He with a superfluid $^4$He overlayer. With nominal surface $^4$He coverage in the range 68 to 139 $\mu$mol/m$^2$, we always detect the same specularity, consistent with previous work[32], which found evidence for surface scattering close to specular for surface film coverages in excess of 60 $\mu$mol/m$^2$.

**Theoretical calculation of gap suppression** is made using quasiclassical theory with the random S-matrix scattering model (Supplementary Note 3).

## Author contributions:

Experimental work was carried out by P.J.H and L.V.L with contributions from A.C. The nanofabricated cells were prepared and assembled by N.Z, X.R and A.C. X.R carried out the FEM simulations of the cell. The analysis and presentation was carried out by P.J.H, L.V.L and A.V with contributions from J.S. A.V performed calculations of gap profile and superfluid transition temperature. A.V developed the theory of magnetic scattering with contributions from P.S. The work at Cornell was supervised by J.M.P, and the work in London was supervised by J.S, A.C and L.V.L, who had the leading roles in formulating the research. P.J.H, J.S and A.V had leading roles in writing the paper, with contributions from all authors.

**Competing financial interests:** The authors declare no competing financial interests.

**Materials & Correspondence:** Requests should be addressed to petri.heikkinen@rhul.ac.uk and j.saunders@rhul.ac.uk.

## Figure 1.

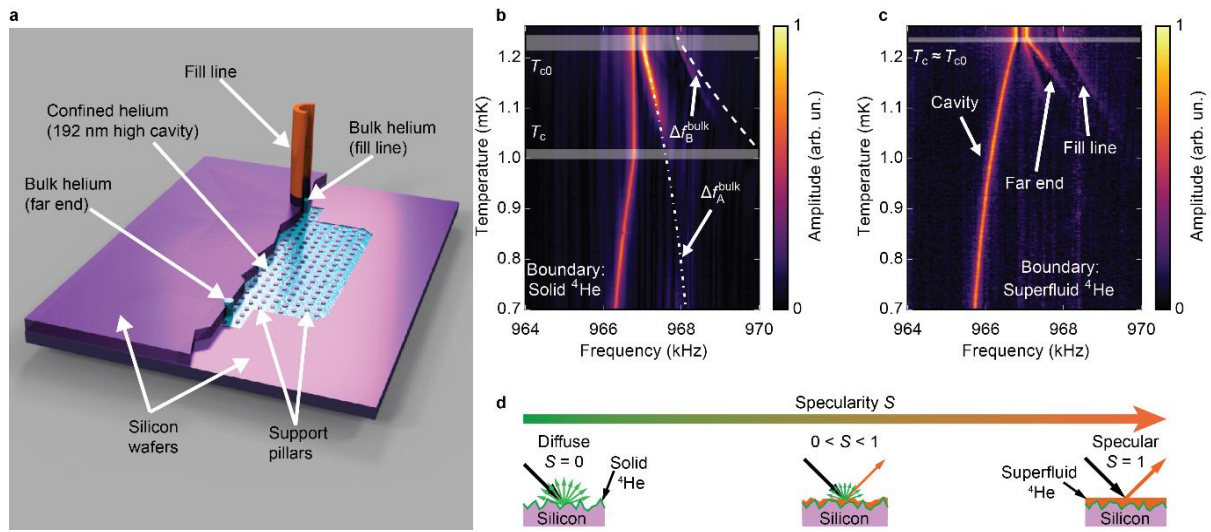

**Figure.1** Experimental cell confining $^3$He. **a**. Nanofabricated sample cell, cut away to show cavity in lower silicon wafer, bonded to upper wafer. The support posts shown maintain cavity height $D$ under different liquid pressures. The cavity is filled through a fill line via a sintered heat exchanger and cooled through the column of $^3$He within it. Small volumes of bulk liquid at each end of the cavity provide markers for the bulk superfluid transition $T_{c0}$, and eliminate errors due to temperature gradients. The NMR coil set around the sample is shown in Supplementary Fig. 1. Suitable small magnetic field gradients are used to resolve the NMR response of different regions of the cell, see Methods. **b, c**. NMR signatures of superfluid transition in cavity and bulk markers, for two different surface boundary conditions, at $^3$He pressure of 2.46 bar. $^3$He-A in the cavity shows a negative frequency shift whereas the bulk markers show positive frequency shift (Supplementary Note 2). The $T_c$ suppression observed with a surface boundary layer of solid $^4$He is eliminated by the addition of $^4$He to create a superfluid $^4$He film at the surface. **d**. Schematic illustration of the tuning of surface scattering conditions, parameterised by specularity coefficient $S$.

**Figure 2.**

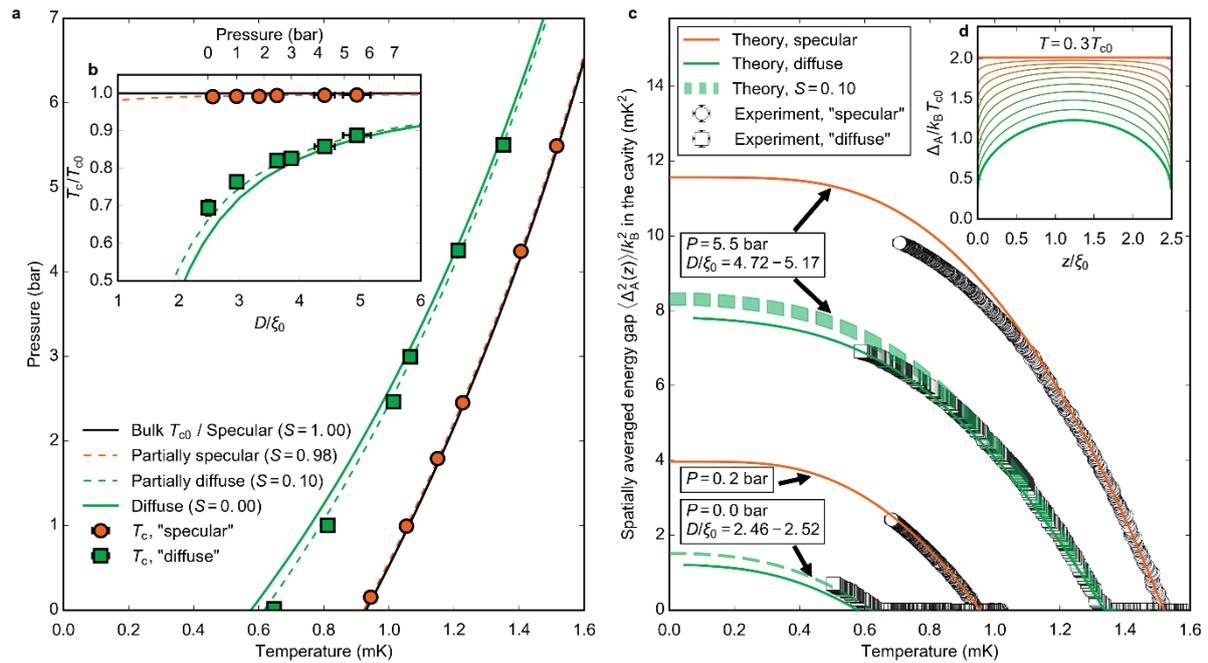

**Figure 2.** Suppression of superfluid transition temperature and superfluid gap for different surface scattering conditions. **a.** Measured pressure dependence of $T_c$ for close to diffuse and close to specular boundary conditions. Full lines show predicted $T_c$ for diffuse and fully specular boundaries, dashed lines are best fits yielding $S = 0.1$ and $S = 0.98$. **b.** Suppression of $T_c$ for diffuse boundary steeply increases with confinement. Suppression of $T_c$ for specular boundary is essentially eliminated. **c.** Spatial average of energy gap in cavity at two pressures (0.0, 5.5 bar), inferred from measured frequency shift (Supplementary Note 1), for these two surface scattering conditions. At zero bar the "diffuse" experiment agrees best with theory for $S = 0.1$. All theoretical curves include strong coupling corrections valid near $T_c$. The emergent discrepancy between theory and experiment at lower temperatures at 5.5 bar, for both scattering conditions, is in agreement with the expected temperature dependence of strong coupling corrections to the gap (Supplementary Note 4). The width of the theoretical curves for $S = 0.1$ accounts for errors associated with the weak pressure dependence of cavity height (Methods). **d.** The calculated gap profile at zero pressure for specularities between 0 and 1 in intervals of 0.1.

**Figure 3.**

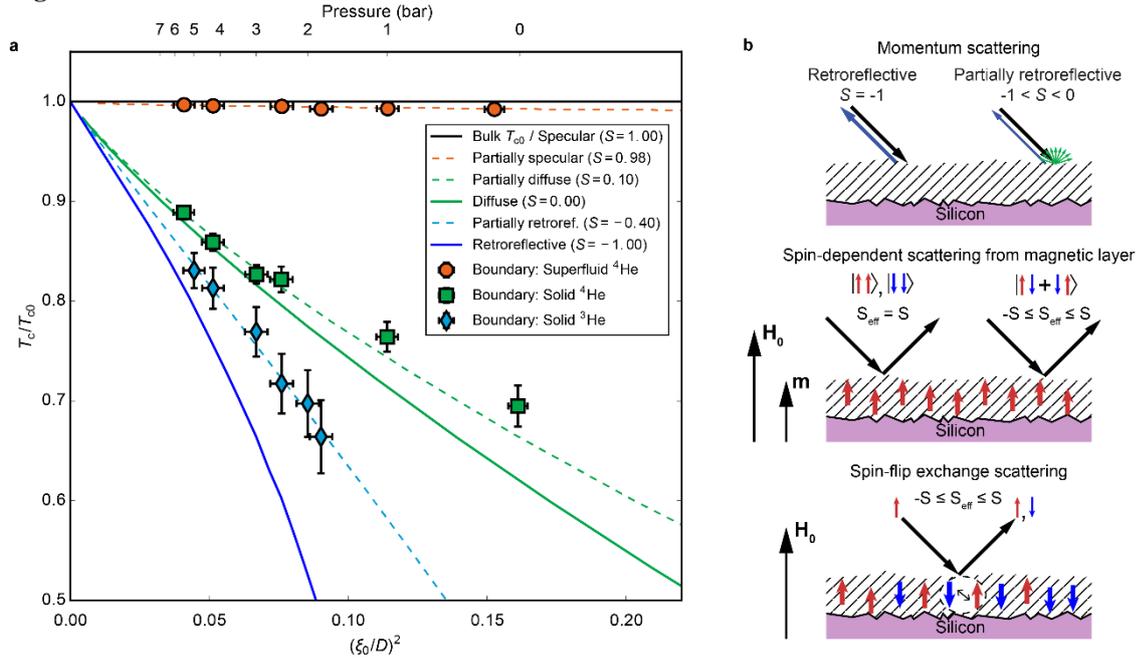

**Figure 3.a.** Increased suppression of superfluid transition temperature in presence of a magnetic solid ³He surface boundary layer. Plot compares suppression of $T_c$ in cavity relative to that of bulk liquid as a function of square of the inverse effective confinement, for solid ³He boundary (diamond), solid ⁴He boundary (square) and superfluid ⁴He boundary (circles). Full lines show: maximal pair-breaking retro-reflection ($S = -1$); diffuse ($S = 0$); fully specular ($S = 1$). Dashed lines show best fits to the data: $S = -0.4, 0.1, 0.98$. For the solid ³He boundary, $T_c$ is identified from onset of superfluid frequency shift after correcting for background frequency shift arising from magnetic solid layer (Supplementary note 5). **b.** Three candidate scattering mechanisms for negative effective specularity (see also Supplementary Note 6): retroreflection (ruled out by normal state measurements); spin-dependent pair breaking on scattering from a magnetically polarized layer (absent for the relative orientation of surface layer spin polarization **m**, surface normal and spin orientation of A-phase pairs in our set-up); spin-flip exchange scattering (spin-polarization of surface layer can be zero) Here the effective specularity $S_{eff}$ is a parameter characterising combined magnetic and momentum scattering, bounded by the specularity $S$ that would arise from momentum scattering alone.

**Figure 4.**

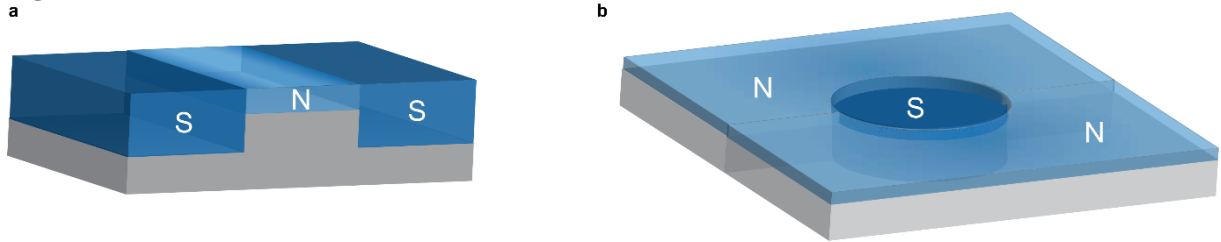

**Figure 4.** Topological mesoscopic superfluidity. Confinement tunes $^3$He into different material phases, enabling hybrid structures. **a.** An SNS junction, where spatial modulation of cavity height defines SN interfaces. **b.** Circular region of higher cavity height defines an isolated mesa of superfluid, cooled through normal liquid in a more confined region.



\clearpage

\maketitle

\begin{figure*}
	\includegraphics[width=\textwidth]{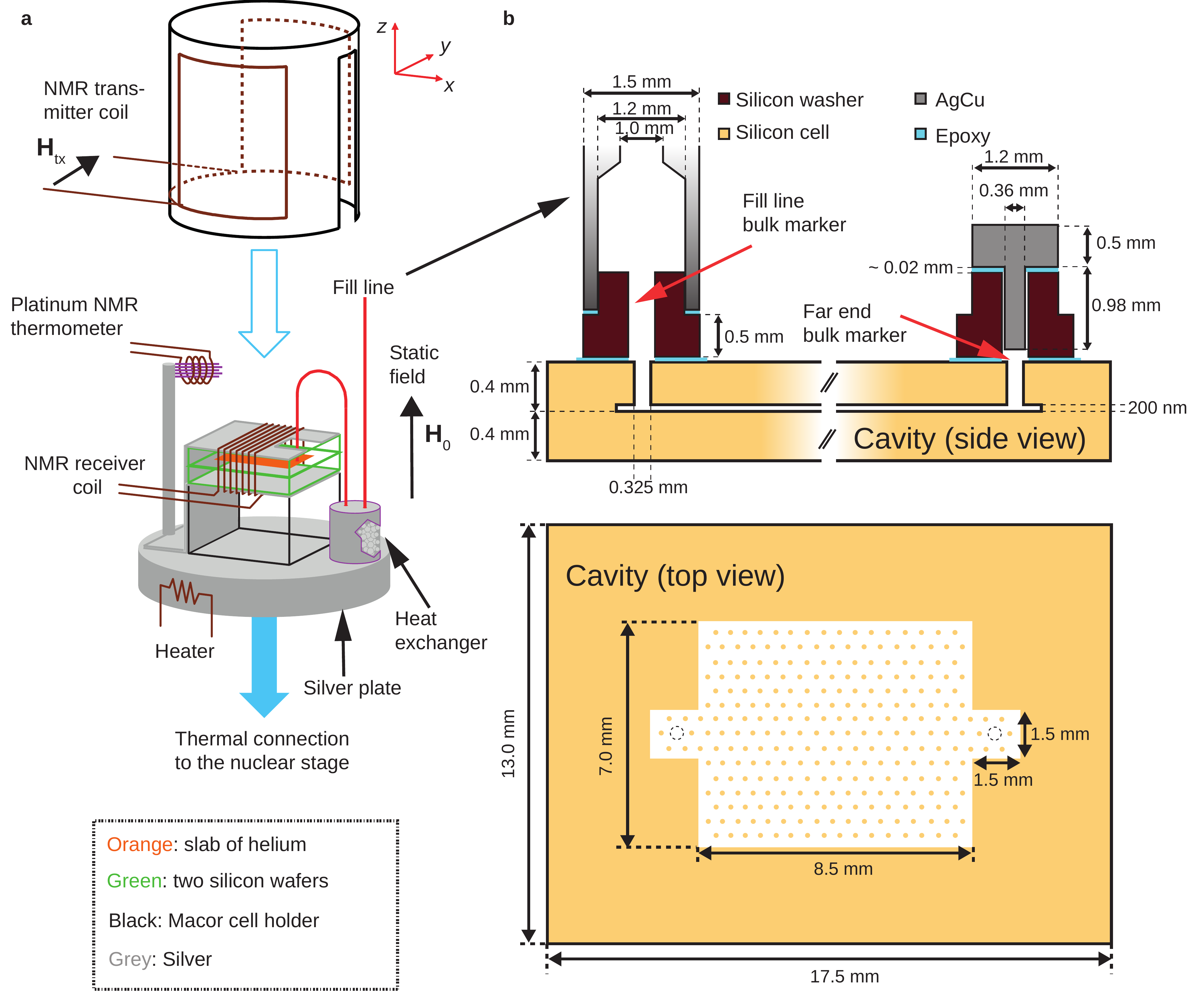}
	\caption{\label{fig:full_setup} \textbf{Setup for NMR measurements.} \textbf{a,} Schematics of the NMR setup. Silicon-silicon sample container sits on a silver plate thermally connected to the copper nuclear stage of the cryostat. Silver foils glued on both sides of the silicon structure and screwed on the silver plate help to thermalise the sample container. Nanofabricated slab-shaped cavity is filled with helium through a sintered silver heat exchanger having surface area 8\,m$^2$. $^3$He in the cavity is cooled via the $^3$He column in the fill line. Temperature of the silver plate is measured using a platinum NMR thermometer. Main NMR field $\mathbf{H}_0$ is created with a solenoid coil located inside the 4\,K bath far from the sample region. Saddle-shaped transmitter coil used to create the NMR pulse field $\mathbf{H}_{\mathrm{tx}}$ is wound on sides of a Macor holder sliding around the sample region on the silver plate. The precessing sample magnetisation is measured with a receiver coil wound tightly around the sample container. \textbf{b,} Dimensions of the sample container. Most of the volume in the fill line is shielded from NMR measurements by the metallic fill line. The small unshielded part on the bottom end as well as the small compartment on the far end of the cavity have volumes of the same order as the cavity to result in three comparable peaks in NMR spectra. Triangular lattice of 100\,\textmu m diameter pillars separated by 500\,\textmu m reduces the distortion of cavity due to liquid overpressure.}
\end{figure*}

\begin{figure*}
	\includegraphics[width=0.7\textwidth]{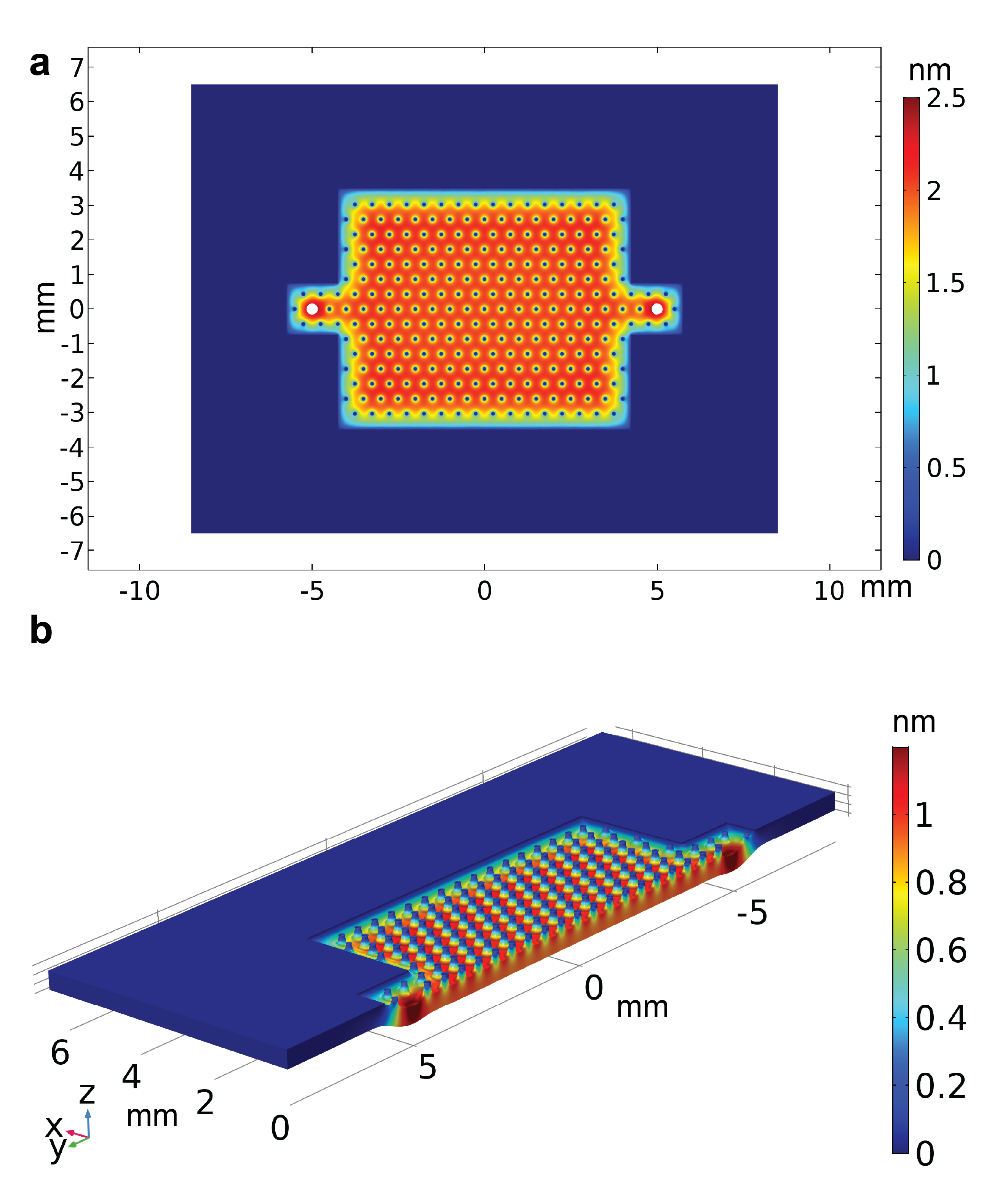}
	\caption{\label{fig:FEM_bowing} \textbf{FEM simulation of distortion of the cavity due to overpressure.} \textbf{a,} Change in the cavity height due to 1\,bar internal pressure, as simulated with finite element method (FEM), using COMSOL Multiphysics, increases sharply when moving from the cavity edges towards the centre. Maximal simulated distortion with triangular lattice of support pillars in the centre region is 2.1\,nm/bar. We estimate a 0.5\,nm/bar uncertainty in the simulation and thus use higher value 2.6\,nm/bar for the maximal cavity height distortion in analysis. \textbf{b,} Three-dimensional illustration of the internal cavity distortion. The effect has been greatly magnified here for clarity. We use the same material parameters for silicon as in Ref.~\cite{Zhelev_RevSciInst2018}: density $\rho^{}_{\mathrm{Si}} = 2329\,$kg/m$^3$, the Young's modulus $E_{\mathrm{Si}} = 170\,$GPa, and the Poisson's ratio $\nu^{}_{\mathrm{Si}} = 0.28$.}
\end{figure*}

\begin{itemize}
	\setlength{\itemindent}{-2em}
	\item[] \textbf{CONTENTS}
	\item[] \textbf{Supplementary Fig.~\ref{fig:full_setup}:} Setup for NMR measurements
	\item[] \textbf{Supplementary Fig.~\ref{fig:FEM_bowing}:} FEM simulation of distortion of the cavity due to overpressure	
	\item[] \textbf{Supplementary Note~\ref{sec:fshift_vs_gap}:} Connection between frequency shifts and energy gap
	\begin{itemize}
		\item[] \textbf{Supplementary Fig.~\ref{fig:IS_example}:} Determination of initial slopes in the cavity
		\item[] \textbf{Supplementary Fig.~\ref{fig:IS_comparison}:} Comparison of measured initial slopes to earlier results and theory
	\end{itemize}
	\item[] \textbf{Supplementary Note~\ref{sec:bulk_markers}:} Bulk marker frequency shifts and temperature gradient across cavity
	\begin{itemize}
		\item[] \textbf{Supplementary Fig.~\ref{fig:fshift_comparison}:} Bulk marker frequency shifts
		\item[] \textbf{Supplementary Fig.~\ref{fig:T_gradient}:} Temperature gradient and temperature hysteresis across the cavity
	\end{itemize}
	\item[] \textbf{Supplementary Note~\ref{sec:theory_OP}:} Theoretical calculation of the order parameter in $^3$He slab geometry
	\begin{itemize}
		\item[] \textbf{Supplementary Fig.~\ref{fig:gap_prof}:} Spatial dependence of energy gap in slab-shaped cavity
		\item[] \textbf{Supplementary Fig.~\ref{fig:DOS_vs_S}:} Density of states of $^3$He-A at the surface and in the middle of the cavity
	\end{itemize}
	\item[] \textbf{Supplementary Note~\ref{sec:strong-coupling}:} Strong-coupling corrections
	\begin{itemize}
		\item[] \textbf{Supplementary Fig.~\ref{fig:strong-coupling}:} Strong-coupling corrections to bulk energy gap
	\end{itemize}
	\item[] \textbf{Supplementary Note~\ref{sec:solid3He}:} Determination of $T_{\mathrm{c}}$ in presence of solid $^3$He surface boundary layer
	\begin{itemize}
		\item[] \textbf{Supplementary Note 5.1:} Sample magnetisation
		\item[] \textbf{Supplementary Fig.~\ref{fig:CW}:} Magnetisation with solid layer of $^3$He on the cavity walls
		\item[] \textbf{Supplementary Note 5.2:} Extraction of superfluid transition temperature
		\item[] \textbf{Supplementary Fig.~\ref{fig:solid3He_Tc}:} Superfluid frequency shift in the cavity with solid $^3$He on the walls
	\end{itemize}
	\item[] \textbf{Supplementary Note~\ref{sec:magnetic_scattering}:} Pair breaking at the surface
	\begin{itemize}
		\item[] \textbf{Supplementary Note 6.1:} Momentum scattering
		\item[] \textbf{Supplementary Note 6.2:} Magnetic scattering from polarized surface
		\item[] \textbf{Supplementary Note 6.3:} Quantum spin scattering
		\item[] \textbf{Supplementary Fig.~\ref{fig:spinTc}:} Suppression of $T_{\mathrm{c}}$ by magnetic scattering on quantum spins
	\end{itemize}
	\item[] \textbf{Supplementary Note~\ref{sec:T_correction}:} Temperature correction
	\begin{itemize}
		\item[] \textbf{Supplementary Fig.~\ref{fig:NMR_calib}:} Calibration of heating caused by NMR pulses
		\item[] \textbf{Supplementary Fig.~\ref{fig:NMR_corr}:} Model parameters used in temperature correction
	\end{itemize}
	\item[] \textbf{Supplementary References [1]--[48]}
\end{itemize}

\clearpage

\section{Connection between frequency shifts and energy gap}
\label{sec:fshift_vs_gap}

Here we restrict the discussion to linear spin dynamics relevant to our NMR experiments with small spin tipping angles $\beta$. The following discussion covers the observed frequency shift in the cavity; the frequency shifts observed in the bulk marker volumes are discussed in Supplementary Note~\ref{sec:bulk_markers}.

In normal $^3$He the NMR precession occurs at the Larmor frequency $f_{\mathrm{L}} = \gamma H_0/2\pi$, where $\gamma$ is the gyromagnetic ratio of $^3$He. Frequency shifts $\Delta f_{\mathrm{sf}} = f - f_{\mathrm{L}}$ of different superfluid phases below $T_{\mathrm{c}}$ are determined by the curvature of dipole energy as a function of rotations in spin space~\cite{VW}. Both positive and negative frequency shifts are possible depending on whether the dipole energy is at minimum or at maximum before the spin-tipping NMR pulse, respectively. In $^3$He-A the order parameter, and thus the magnitude of the dipole energy, is parametrised with two unit vectors $\hat{\mathbf{l}}$ and $\hat{\mathbf{d}}$ exhibiting long-range order. Here $\hat{\mathbf{l}} \parallel \mathbf{L}$ describes the orientation of the orbital angular momentum of all the Cooper pairs, and the order-parameter vector $\mathbf{d}$ points in the direction of zero spin projection: $\hat{\mathbf{d}} \perp \mathbf{S}$. In general, the directions of both vectors are determined by a competition between orienting effects including external magnetic field, boundaries of sample, dipole energy, and superfluid flow. Applied magnetic field $H_0$ larger than dipolar field $H_{\mathrm{D}} \sim 5\,$mT locks $\hat{\mathbf{d}}\perp \mathbf{H}_0$, whereas the dipole energy is minimised when $\hat{\mathbf{l}} \parallel \hat{\mathbf{d}}$. Any surface introduces strong boundary condition $\hat{\mathbf{l}} \parallel \hat{\mathbf{s}}$ where $\hat{\mathbf{s}}$ is normal to the surface. The length scale over which boundaries affect the order parameter is determined by the dipolar length $\xi_{\mathrm{D}} \sim 10\,$\textmu m~\cite{VW}.

We have $\mathbf{H}_0 \parallel \hat{\mathbf{z}}$ and $\hat{\mathbf{s}} \parallel \hat{\mathbf{z}}$. The magnetic field defines the orientation $\hat{\mathbf{d}} \perp \hat{\mathbf{z}}$ and the small height of the cavity ($D\ll \xi_{\mathrm{D}}$) forces $\hat{\mathbf{l}} \parallel \hat{\mathbf{z}}$ everywhere in it. This dipole-unlocked configuration maximises the dipole energy, resulting in a negative NMR frequency shift in the A phase~\cite{Ahonen_1976_JLTP,RHUL_Science}:
\begin{equation}
\label{eq:df_A_bulk}
f^2-f_{\mathrm{L}}^2 = -\frac{\gamma^2 \lambda_{\mathrm{D}} N_{\mathrm{F}}}{5\pi^2 \chi^{}_{\mathrm{N}}}\Delta_{\mathrm{A}}^2 = -\left(\frac{\Omega_{\mathrm{A}}}{2\pi}\right)^2.
\end{equation}
Here $\Omega_{\mathrm{A}}$ is the A-phase Leggett frequency (the longitudinal resonance frequency) depending on both the temperature and the pressure, and energy gap $\Delta_{\mathrm{A}}$ refers to the maximum A-phase energy gap in the momentum space at $\hat{\mathbf{p}}\perp \hat{\mathbf{l}}$. Here $\lambda_{\mathrm{D}} \sim 10^{-6}$ sets the relative scale of the dipole energy, $N_{\mathrm{F}}$ is the density of states at the Fermi level, and $\chi^{}_{\mathrm{N}}$ is the normal state spin susceptibility equalling the susceptibility of equal-spin-paired $^3$He-A. All pre-factors of $\Delta_{\mathrm{A}}^2$ depend only on the normal state properties of $^3$He independent of the level of confinement or boundary condition. Thus, all the dependences of the cavity frequency shift at constant pressure are fully defined by the energy gap.

In a cavity with non-specular quasiparticle scattering at the walls, the gap acquires a spatially inhomogeneous suppression $\Delta_{\mathrm{A}} (z)$ as seen in Supplementary Fig.~\ref{fig:gap_prof}. However, the NMR precession in a confinement volume where $D \ll \xi_{\mathrm{D}}$ is uniform with the frequency shift~\cite{Dmitriev_2010, RHUL_Science}
\begin{equation}
\label{eq:fshift_vs_mean_gap}
\left|f^2-f_{\mathrm{L}}^2\right| = \frac{\gamma^2 \lambda_{\mathrm{D}} N_{\mathrm{F}}}{5\pi^2 \chi^{}_{\mathrm{N}}}\left<\Delta_{\mathrm{A}}^2(z)\right>,
\end{equation}
where $\left<\Delta_{\mathrm{A}}^2(z)\right>$ refers to the spatially averaged value of the squared energy gap, the experimental determination of which we now discuss. 

For compactness we write $\left| f^2 - f_{\mathrm{L}}^2 \right| = \zeta \left< \Delta_{\mathrm{A}}^2(z) \right>$, where $\zeta =  \frac{\gamma^2 \lambda_{\mathrm{D}} N_{\mathrm{F}}}{5\pi^2 \chi^{}_{\mathrm{N}}}$ is a pressure-dependent, temperature-independent constant. In bulk in the Ginzburg-Landau (G-L) regime, near bulk transition temperature $T_{\mathrm{c0}}$, $\Delta_{\mathrm{A}}^2 \propto \left(1 - T/T_{\mathrm{c0}}\right)$. Therefore, in this regime, we define $\mathrm{IS}_{\Delta}^{\mathrm{bulk}}$ and $\mathrm{IS}^{\mathrm{bulk}}$ by
\begin{equation}
\left| f^2 - f_{\mathrm{L}}^2 \right| = \zeta \mathrm{IS}_{\Delta}^{\mathrm{bulk}} \left(1 - \frac{T}{T_{\mathrm{c0}}}\right) = \mathrm{IS}^{\mathrm{bulk}} \left( 1 - \frac{T}{T_{\mathrm{c0}}} \right).
\end{equation}
Now clearly $\zeta = \mathrm{IS}^{\mathrm{bulk}} / \mathrm{IS}_{\Delta}^{\mathrm{bulk}}$.

For our ``specular'' ($S=0.98$) surface, which shows essentially no $T_{\mathrm{c}}$ suppression, the measured frequency shift corresponds to the spatially uniform and unsuppressed bulk gap $\Delta_{\mathrm{A}}$, Supplementary Eq.~(\ref{eq:df_A_bulk}). Thus, we have for the precession frequency in the cavity
\begin{equation}
\left|f^2 - f_{\mathrm{L}}^2\right| = \mathrm{IS}_{\mathrm{slab}}^{\mathrm{'spec'}} \left(1-\frac{T}{T_{\mathrm{c}}}\right),
\end{equation}
where $T_{\mathrm{c}} = T_{\mathrm{c0}}$ to a good approximation. Here $\mathrm{IS}_{\mathrm{slab}}^{\mathrm{'spec'}} \approx \mathrm{IS}^{\mathrm{bulk}}$ is determined as a slope of a linear fit between the cavity frequency shift and temperature close enough to $T_{\mathrm{c}}$ where the dependence is expected to be nearly linear~\cite{RHUL_Science}:
\begin{equation}
\mathrm{IS}_{\mathrm{slab}}^{\mathrm{'spec'}} = \frac{\partial \left|f^2 - f_{\mathrm{L}}^2\right|}{\partial\left(1-T/T_{\mathrm{c}}\right)},\hspace{1cm} \textrm{averaged over } 0.90T_{\mathrm{c}} < T < T_{\mathrm{c}}.
\end{equation}
Similarly, $\mathrm{IS}_{\Delta}^{\mathrm{bulk}}$ is determined from a linear fit to the calculated bulk gap over equivalent temperature range:
\begin{equation}
\label{eq:IS_Delta}
\mathrm{IS}_{\Delta}^{\mathrm{bulk}} = \frac{\partial\Delta_{\mathrm{A}}^2}{\partial\left(1-T/T_{\mathrm{c0}}\right)},\hspace{1cm} \textrm{averaged over } 0.90T_{\mathrm{c0}} < T < T_{\mathrm{c0}}.
\end{equation}
The choice of this temperature range is a suitable compromise between precision and accuracy and justifies calling the proportionality constants $\mathrm{IS}$ as initial slopes, see Supplementary Fig.~\ref{fig:IS_example}.

Determination of constant $\zeta$ using the ratio of the two abovementioned initial slopes provides a high degree of cancellation of the systematic error arising from the choice of temperature range. This is true for both weak-coupling and strong-coupling models in the G-L regime (Supplementary Note~\ref{sec:strong-coupling}). Now we get
\begin{equation}
\label{eq:fshift_to_gap}
\left<\Delta_{\mathrm{A}}^2(z)\right> = \frac{\mathrm{IS}_{\Delta}^{\mathrm{bulk}}}{\mathrm{IS}_{\mathrm{slab}}^{\mathrm{'spec'}}} \left|f^2 - f_{\mathrm{L}}^2\right|.
\end{equation}
This expression determines the average gap suppression for arbitrary surface scattering at all temperatures. The procedure described here eliminates any systematic errors that might arise from the use of literature values of the bulk frequency shift. To compare the experimental and theory-based specular and non-specular initial slopes we can use the dependence:
\begin{equation}
\label{eq:IS_spec_to_diff}
\frac{\mathrm{IS}_{\mathrm{slab}}^{\mathrm{'diff'}}}{\mathrm{IS}_{\mathrm{slab}}^{\mathrm{'spec'}}} = \frac{\mathrm{IS}_{\Delta}^{\mathrm{'diff'}}}{\mathrm{IS}_{\Delta}^{\mathrm{bulk}}},
\end{equation}
where the superscript 'diff' refers to any non-specular boundary condition. This dependence holds as long as the suppressed gap $\left< \Delta_{\mathrm{A}}^2(z) \right> = \mathrm{IS}_{\Delta}^{\mathrm{'diff'}} \left(1 - T/T_{\mathrm{c}}\right)$.

\begin{figure*}
	\includegraphics[width=\textwidth]{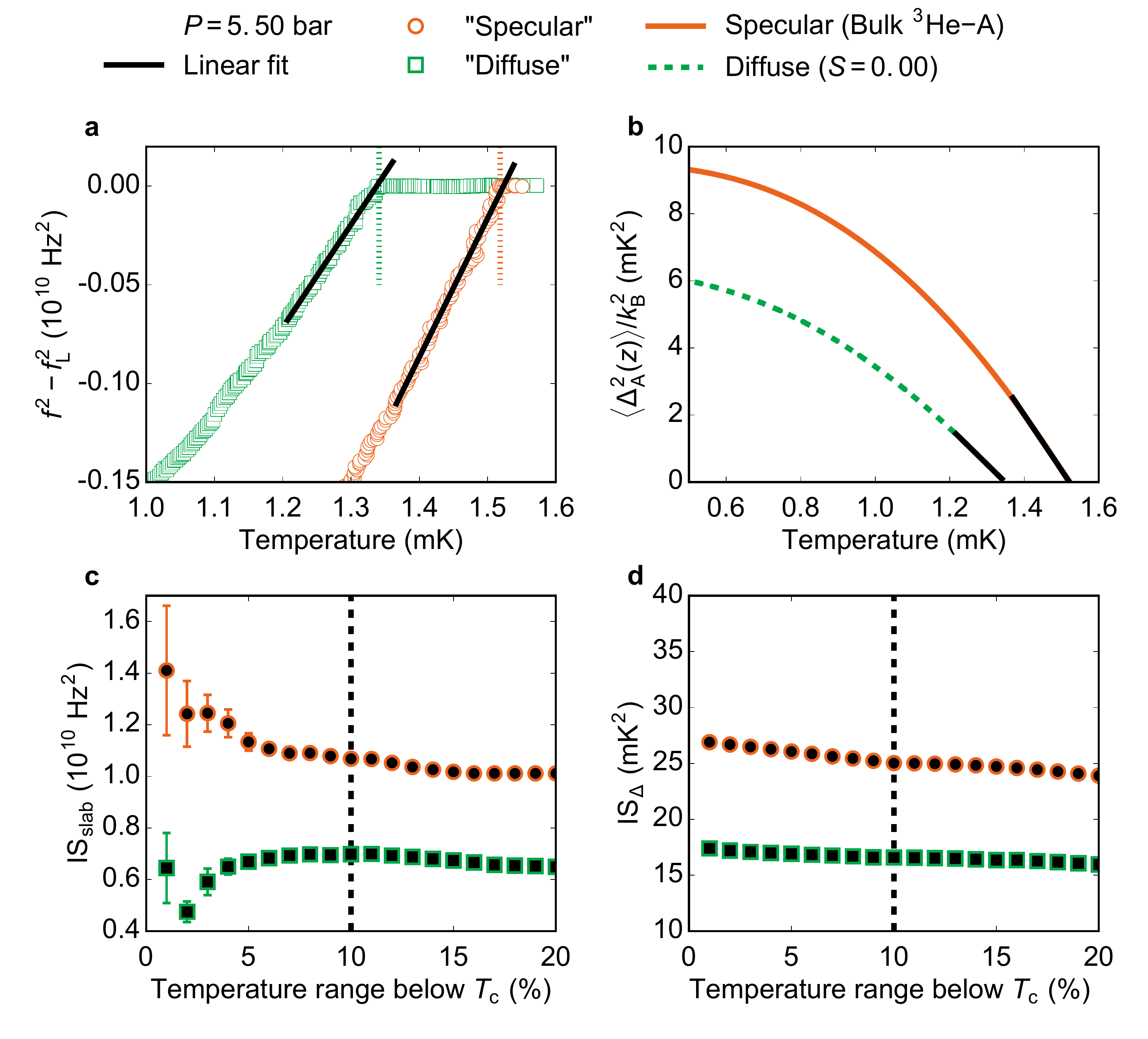}
	\caption{\label{fig:IS_example} \textbf{Determination of initial slopes in the cavity.} \textbf{a,} Measured frequency shift $f^2 - f_{\mathrm{L}}^2$ against temperature with ``specular'' and ``diffuse'' boundary conditions at $P=5.50\,$bar, showing the linear fits over the range $0.90T_{\mathrm{c}} < T < T_{\mathrm{c}}$ used to determine the experimental initial slopes $\mathrm{IS}_{\mathrm{slab}}$. Vertical dotted lines indicate the measured $T_{\mathrm{c}}$. \textbf{b,} Calculated spatially averaged values of the squared quasiclassical weak-coupling energy gap with specular and diffuse boundary conditions. Linear fits over the range $0.90T_{\mathrm{c}} < T < T_{\mathrm{c}}$ are used to determine the theory-based initial slopes $\mathrm{IS}_{\Delta}$. \textbf{c} Initial slopes $\mathrm{IS}_{\mathrm{slab}}$ as a function of temperature range of the linear fit below measured $T_{\mathrm{c}}$. Increased uncertainty limits and scatter approaching $T_{\mathrm{c}}$ arise from limited amount of data combined with noise, temperature gradient across the cavity (Supplementary Fig.~\ref{fig:T_gradient}), and potential rounding of $T_{\mathrm{c}}$ due to cavity height distortion (Supplementary Fig.~\ref{fig:FEM_bowing}). \textbf{d,} Initial slopes $\mathrm{IS}_\Delta$ as a function of temperature range of the linear fit below calculated $T_{\mathrm{c}}$. We estimate the systematic error between the actual initial slope and the slope determined over the range $0.90T_{\mathrm{c}} < T < T_{\mathrm{c}}$ to be 8\%. The range chosen for conversion between the frequency shift and the energy gap, Supplementary Eq.~(\ref{eq:fshift_to_gap}), is marked by vertical dashed line in \textbf{c} and \textbf{d}.}
\end{figure*}

\begin{figure*}
	\includegraphics[width=\textwidth]{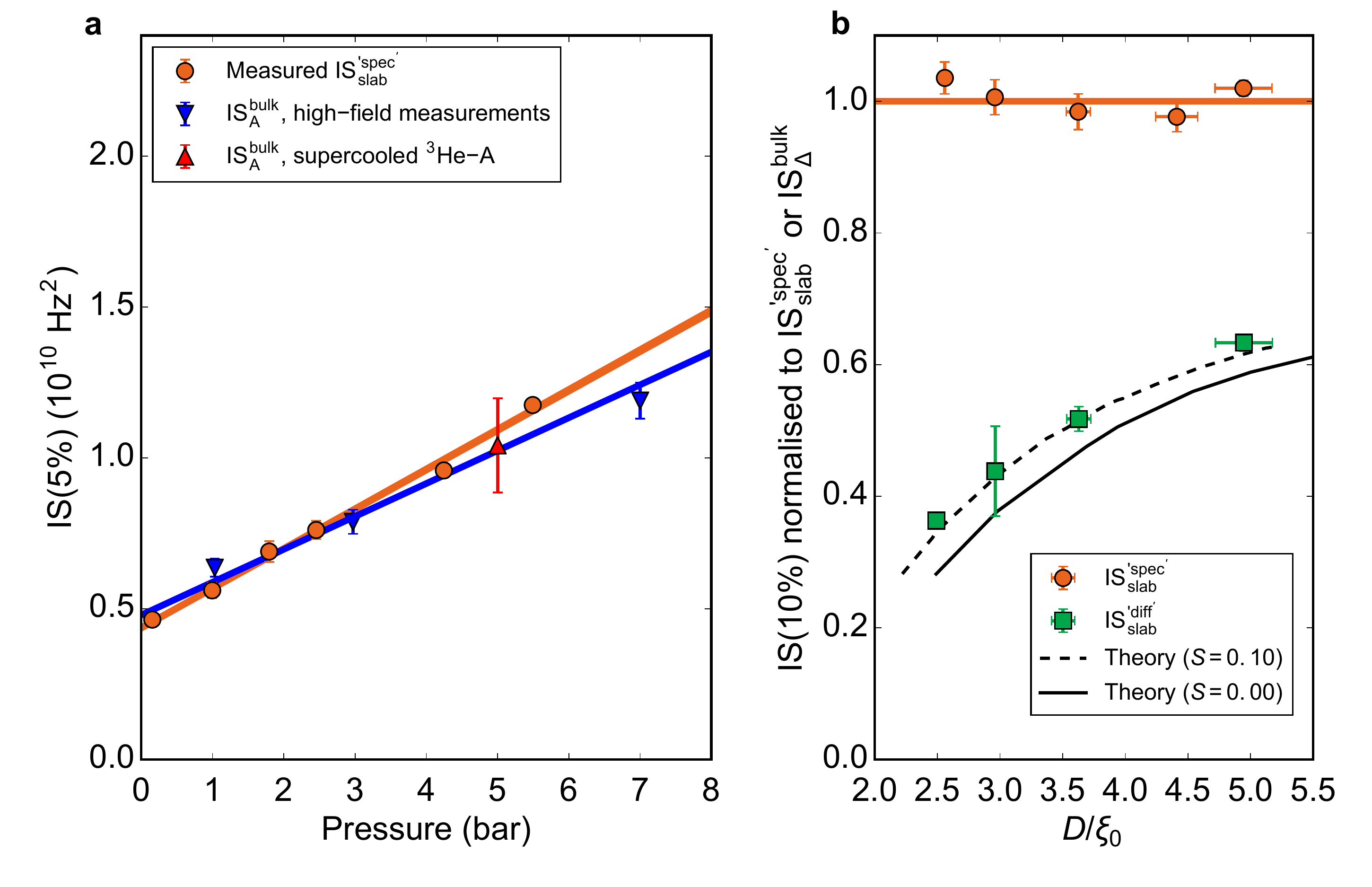}
	\caption{\label{fig:IS_comparison} \textbf{Comparison of measured initial slopes to earlier results and theory.} \textbf{a,} Measured $\mathrm{IS}_{\mathrm{slab}}^{\mathrm{'spec'}}$ (5\% range below $T_{\mathrm{c}}$) in cavity with ``specular'' boundary condition show a linear low-pressure dependence (orange line) similar to the previous experimental data. Blue line is a linear fit given in Ref.~\cite{Zimmerman_2018} based on $\mathrm{IS}_{\mathrm{A}}^{\mathrm{bulk}}$ measured using transverse NMR frequency shifts in bulk $^3$He-A at pressures below polycritical point (21.22\,bar) at high magnetic fields (blue downward-pointing triangles)~\cite{Rand_thesis,Rand_Physica}. In Ref.~\cite{bulk_A_IS} (red upward-pointing triangle) the frequency shifts were measured by supercooling bulk A phase significantly below its stable field-pressure configuration in the phase diagram. They determined initial slopes over the range $0.90T_{\mathrm{c0}} < T < T_{\mathrm{c0}}$, so here we scale their data into the 5\% range, using the estimated difference in the systematic errors (see Supplementary Fig.~\ref{fig:IS_example}d). \textbf{b,} Initial slopes for ``diffuse'' scattering normalised to measured ``specular'' initial slopes are compared to quasiclassical weak-coupling theory giving $\mathrm{IS}_{\Delta}^{\mathrm{'diff'}} / \mathrm{IS}_{\Delta}^{\mathrm{bulk}}$ (see Supplementary Eq.~(\ref{eq:IS_spec_to_diff})) in the range $0.90T_{\mathrm{c}} < T < T_{\mathrm{c}}$. The data agree best with $S=0.10$ curve.}
\end{figure*}

Supplementary Fig.~\ref{fig:IS_comparison}a shows the comparison between the measured values of the initial slopes and the earlier experimental values from Refs.~\cite{Rand_thesis,Rand_Physica,bulk_A_IS} in the pressure range covered in the current experiments. Since the previous experiment with largest pressure overlap with us used approximately 5\% range below $T_{\mathrm{c0}}$ ($0.95T_{\mathrm{c0}} < T < T_{\mathrm{c0}}$) to determine the bulk initial slopes~\cite{Rand_thesis,Rand_Physica}, we extract the corresponding values from our ``specular'' measurements for direct comparison instead of using the 10\% range defined above. This eliminates the difference between systematic errors. As expected, the measured initial slopes, $\mathrm{IS}_{\mathrm{slab}}^{\mathrm{'spec'}}$, compare well with the values obtained in bulk $^3$He-A, $\mathrm{IS}_{\mathrm{A}}^{\mathrm{bulk}}$. The linear fit given in Ref.~\cite{Zimmerman_2018} to original data from Refs.~\cite{Rand_thesis, Rand_Physica} is $\mathrm{IS}_{\mathrm{A}}^{\mathrm{bulk}} = \left( 0.479 + 0.109 \frac{1}{\textrm{bar}} P \right) \cdot 10^{10}\,\textrm{Hz}^2$, whereas the fit to current experiments using the equivalent 5\% range is $\mathrm{IS}_{\mathrm{slab}}^{\mathrm{'spec'}} = \left( 0.436 + 0.131 \frac{1}{\textrm{bar}} P \right) \cdot 10^{10}\,\textrm{Hz}^2$. The earlier fit is based on data taken up to 22\,bar, so higher pressure experiments under specular confinement would be required to see whether any significant difference persists or whether the seen difference between the fits is due to scatter in the experimental values. 

The initial slopes determined for ``diffuse'' boundary condition are compared to theory in Supplementary Fig.~\ref{fig:IS_comparison}b. We see that our experiments agree well with specularity $S=0.10$, further confirming that to be the best value of specularity for the boundaries with 32\,\textmu mol/m$^2$ $^4$He preplating.

\clearpage

\section{Bulk marker frequency shifts and temperature gradient across cavity}
\label{sec:bulk_markers}

Helium in the bulk liquid compartments on both ends of the cavity can nucleate either into $^3$He-B or $^3$He-A at bulk superfluid transition temperature $T_{\mathrm{c0}}$. Supercooling of the A phase is more likely in the far-end bulk compartment which is isolated from the rest of the system via nanofluidic cavity. The frequency shift in bulk $^3$He-A, which is at minimum of dipole energy, has the same magnitude but opposite sign compared to frequency shift in the cavity, Supplementary Eq.~(\ref{eq:df_A_bulk}). The bulk B-phase frequency shift is
\begin{equation}
\label{eq:df_B_bulk}
f_{\mathrm{B}}^2 - f_{\mathrm{L}}^2 = \frac{3\gamma^2 \lambda_{\mathrm{D}} N_{\mathrm{F}}}{4\pi^2\chi^{}_{\mathrm{B}}} \Delta_{\mathrm{B}}^2 \sin^2\beta_n =
\left(\frac{\Omega_{\mathrm{B}}}{2\pi}\right)^2 \sin^2\beta_n,
\end{equation}
where $\Omega_{\mathrm{B}}$ is the B-phase Leggett frequency, $\Delta_{\mathrm{B}}$ is the isotropic B-phase energy gap, $\chi^{}_{\mathrm{B}}$ is the temperature-dependent B-phase spin susceptibility, and $\beta_n$ is the angle between $\mathbf{H}_0$ and spin-orbit rotation axis $\hat{\mathbf{n}}$~\cite{VW}. The preferred orientation in bulk is such that $\beta_n = 0$. However, at vertical walls of the bulk marker compartments, dipolar energy forces $\beta_n\approx 63.4^\circ$ leading to $\sin^2\beta_n = 0.8$ and to positive frequency shift from Larmor value detected at superfluid transitions. Since the frequency shifts in our experiments are small compared to the Larmor frequency, we use correspondence $f^2-f_{\mathrm{L}}^2 \approx 2f_{\mathrm{L}}\Delta f$ for a straightforward conversion between calculated energy gap and frequency shift.

Examples of detected frequency shifts against temperature are presented in Supplementary Fig.~\ref{fig:fshift_comparison} and in Fig. 1b,c in the main text. The wall-determined value of B-phase frequency shift is the dominant one in the fill-line bulk marker while the supercooled A phase dominates the far-end bulk marker. Tracking of bulk marker frequency shifts to the lowest temperatures has not been successful, since their amplitudes rapidly decrease below $T_{\mathrm{c0}}$ and vanish into noise at $T \approx 0.9T_{\mathrm{c0}}$. In the B phase a temperature-dependent drop in amplitude is expected due to both decreasing value of $\chi_{\mathrm{B}}$ and increasing value of $\Omega_{\mathrm{B}}$, which together with a smooth bending of $\hat{\mathbf{n}}$ between wall-favoured and field-favoured orientations in a macroscopic compartment results in spectral broadening of the signal~\cite{VW}. However, the detected drop is much faster than expected and neither of these effects concerns the A phase. In the usual experiments, where the cavity signal is of the highest interest, a field gradient along the $z$ axis, used to separate the bulk marker signals from the cavity signal, causes additional broadening, but this effect is independent of temperature. Even with optimized bulk marker signal detection --- after removing the field gradient along the $z$ axis and diminishing the cavity signal by applying a field gradient on the $xy$ plane --- the bulk marker amplitudes drop below detection level soon below $T = 0.9T_{\mathrm{c0}}$. The reason for this remains unknown.

Possible temperature gradient across the cavity is seen as different measured temperatures of the silver plate, $T_{\mathrm{Ag}}$, during the superfluid transitions in the bulk volumes. These two silver-plate temperatures are denoted by $T_{\mathrm{c}}^{\mathrm{fill}}$ and $T_{\mathrm{c}}^{\mathrm{far}}$. After the correction for the thermal gradient across the heat exchanger (Supplementary Note~\ref{sec:T_correction}), $T_{\mathrm{c}}^{\mathrm{fill}} \approx T_{\mathrm{c}}^{\mathrm{G}}$. The small measured temperature gradient as a function of pressure and temperature hysteresis as a function of temperature sweep rate are presented in Supplementary Fig.~\ref{fig:T_gradient}.

\begin{figure*}
	\includegraphics[width=\textwidth]{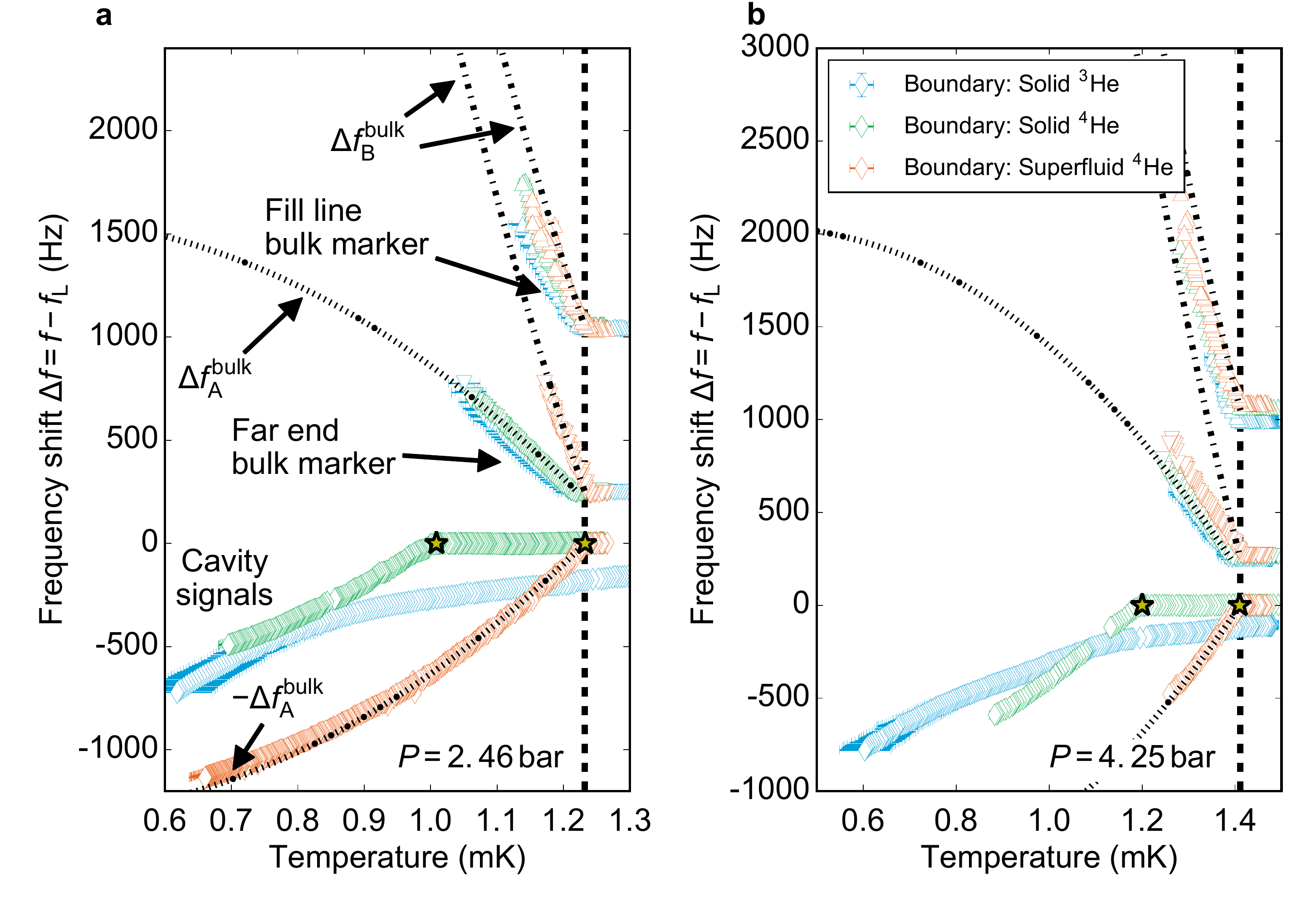}
	\caption{\label{fig:fshift_comparison} \textbf{Bulk marker frequency shifts.} \textbf{a,} Composite of frequency shifts of both fill line and far end bulk markers with cavity frequency shifts at 2.46\,bar. The fill line bulk marker always shows the B phase. The far end bulk marker can remain in the supercooled A phase to the lowest temperatures at which signal is detected. Vertical dashed line indicates the literature value of bulk superfluid transition, $T_{\mathrm{c}}^{\mathrm{G}}$~\cite{Greywall_1986}. Calculated bulk frequency shifts $\Delta f_{\mathrm{A}}^{\mathrm{bulk}}$ (dotted lines) are converted from weak-coupling energy gap $\Delta_{\mathrm{A}}$, using Supplementary Eq.~(\ref{eq:fshift_to_gap}) and measured ``specular'' initial slopes, whereas $\Delta f_{\mathrm{B}}^{\mathrm{bulk}}$ (dashed-dotted lines, wall-determined value, Supplementary Eq.~(\ref{eq:df_B_bulk})) are directly based on energy gap $\Delta_{\mathrm{B}}$ including temperature-dependent trivial strong-coupling corrections~\cite{QuasiClass,Hydrostatic_Thuneberg}, using parameter library in Ref.~\cite{github_Slava_library}. Stars mark the easily-identifiable ``specular'' and ``diffuse'' superfluid transitions in the cavity. \textbf{b,} Equivalent data at 4.25\,bar.}
\end{figure*}

\begin{figure*}
	\includegraphics[width=\textwidth]{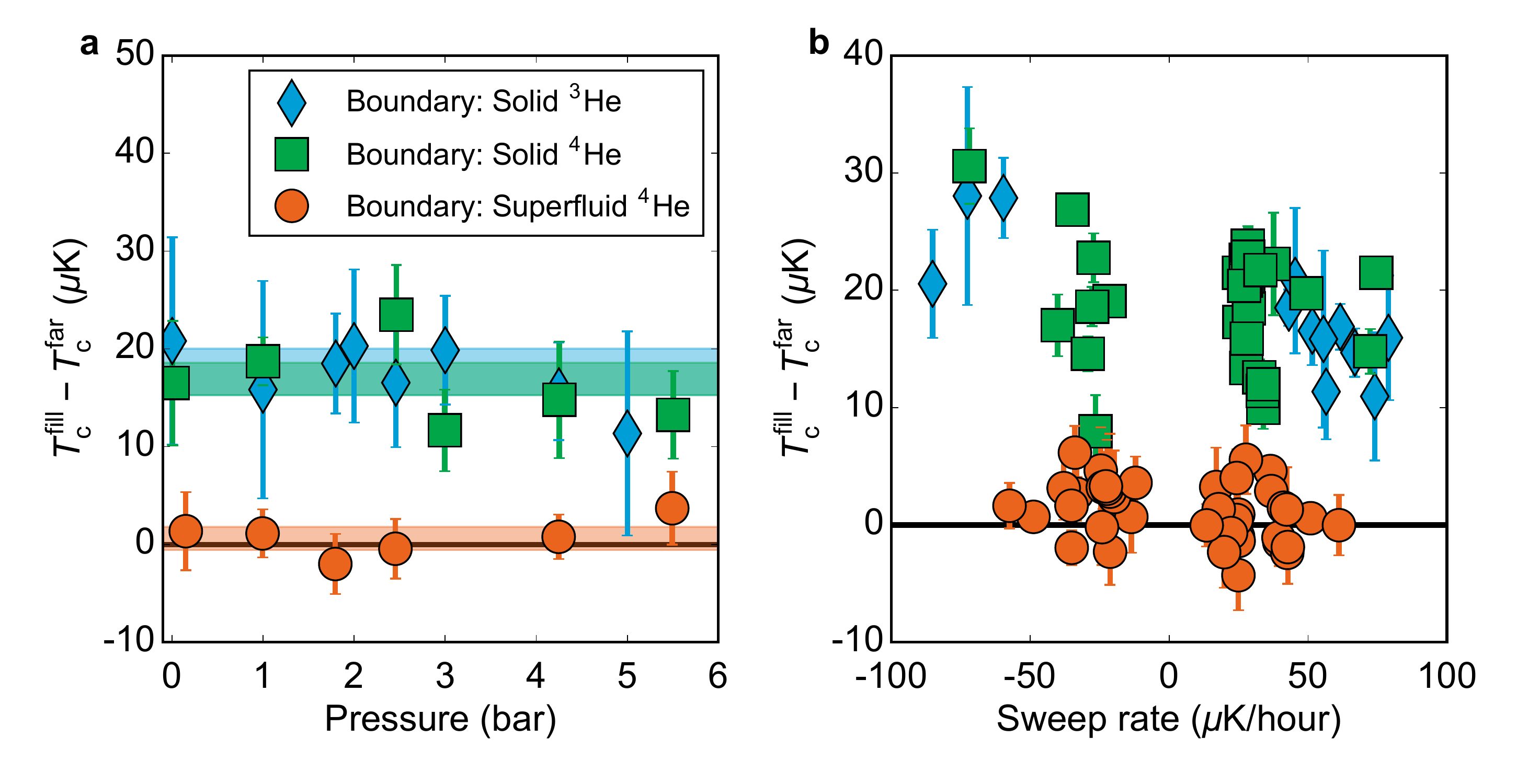}
	\caption{\label{fig:T_gradient} \textbf{Temperature gradient and temperature hysteresis across the cavity.} \textbf{a,} Temperature difference across the cavity at bulk superfluid transition temperature $T_{\mathrm{c0}}$, $\Delta T_{\mathrm{cavity}} = T_{\mathrm{c}}^{\mathrm{fill}} - T_{\mathrm{c}}^{\mathrm{far}}$. With close to specular boundary condition the gradient across the cavity is practically zero at all pressures, whereas with solid $^3$He or $^4$He on the surface the gradient is measured to be around 20\,\textmu K. Each point represents an average of all temperature sweeps conducted at a single pressure. The coloured bands indicate the average value with error for each boundary condition. \textbf{b,} The temperature hysteresis in the cell turns out to be insignificant as characterised by the difference between the measured bulk marker transition temperatures as a function of temperature sweep rate. Most of the measurements have been conducted at sweep rates less than 50\,\textmu K/hour.}
\end{figure*}

\clearpage

\section{Theoretical calculation of the order parameter in $^3$He slab geometry}
\label{sec:theory_OP}

The theoretical procedure to obtain the spatial profile of the order parameter and the corresponding density of states (DOS) in the cavity is based on the quasiclassical theory of superfluid $^3$He~\cite{QuasiClass}. The relevant details for slab geometry can be found in many publications, see for example Refs.~\cite{Vorontsov_2003,Vorontsov_PhilA}.

The multi-component order parameter is determined self-consistently together with the quasiclassical quasiparticle correlator (Green's function) $\widehat{g}(\mathbf{R}, \mathbf{k}; \epsilon_m)$, which describes propagation of quasiparticles with energy $\epsilon_m$ along straight classical trajectories. In cavities thinner than $D \sim 10 \xi_0$, not only the symmetry of the order parameter, but also the nature of the quasiparticle scattering on surfaces determine the behaviour of the quantum condensate. The spectrum of the sub-gap states in the vicinity of the surface is different depending on whether the scattering is specular, diffuse, retroreflecting~\cite{retroref} (maximally pair-breaking in odd-parity superfluid), or a mixture of these, resulting in the different suppression of both the order parameter and the transition temperature. 

The boundary conditions for the quasiclassical propagator at the surface encode the scattering processes with high momentum transfer $\delta k \sim k_{\mathrm{F}}$. We adopt the random $\mathcal{S}$-matrix scattering model~\cite{Nagato_JLTP1996,Nagato_JLTP1998,Nagai_JPSJ2008} with unitary scattering matrix for particle-like excitations, 
\begin{equation}
	\mathcal{S}_{qq'} = -\frac{1-i \eta}{1+i\eta}\Big|_{qq'},
\end{equation}
where $\{ q,q' \}$ are in-plane vectors, and the scattering is from $q'$(in) to $q$(out). For hole-like excitations one has $\tilde{\mathcal{S}}_{q'q} = \mathcal{S}_{qq'}^*$ (final and initial states exchanged), and the entire particle-hole space is covered by 
\begin{equation}
	\widehat{\mathcal{S}} = \left( \begin{array}{cc} \mathcal{S} & 0 \\ 0 & \tilde{\mathcal{S}}^\dag \end{array} \right).
\end{equation}
Random Hermitian matrix $\eta$ has properties 
\begin{equation}
	\overline{\eta^{}_{qq'}} = 0, \qquad \overline{\eta^*_{qq'}\eta^{}_{kk'}} = \kappa(q-q') \delta_{q-q',k-k'},
\end{equation}
where overline refers to statistical average. The correlation cumulant is taken as a constant, $\kappa(q-q') = 2W/\sum_q 1 $~\cite{Nagato_JLTP1998}. Matrix $\eta$ depends on a particular surface scattering picture and for example can be expressed in terms of parameters of the randomly rippled wall model~\cite{Nagato_JLTP1996}.

The $\mathcal{S}$-matrix model allows a continuous description of surface scattering from specular to fully diffuse limit by tuning the $W$-parameter, which we further generalize to include retroreflection by multiplying $\mathcal{S}$ by the overall retro delta function $\delta_{-q,q'}$. The amplitude of the specular reflection is given as
\begin{equation}
	\label{eq:specularity}
	\overline{ \mathcal{S}_{kp} } = - \delta_{kp} \frac{1-\sigma}{1+\sigma} 
\end{equation}
with surface self-energy $\sigma$ determined from self-consistency equation $\sigma = \overline{ \eta \frac{1}{1+\sigma} \eta} = \frac{2W}{1+\sigma}$. This allows one to define `specularity', i.e., the probability of specular scattering, of the surface as $S = \pm (1-\sigma)^2/(1+\sigma)^2$. The negative sign indicates retroreflection. For $W=0$ one has $\sigma=0$ with either completely specular $S=+1$ or retroreflective $S=-1$ surface; for $W=1$ we get $\sigma=1$ and fully diffuse surface $S=0$ with suppressed $T_{\mathrm{c}}(D)$ that exactly follows~\cite{Nagato_JLTP1998} the original result for diffuse $T_{\mathrm{c}}$ suppression given by Kj\"{a}ldman \textit{et al.} (KKR)~\cite{KKR}.

\begin{figure*}
	\includegraphics[width=\textwidth]{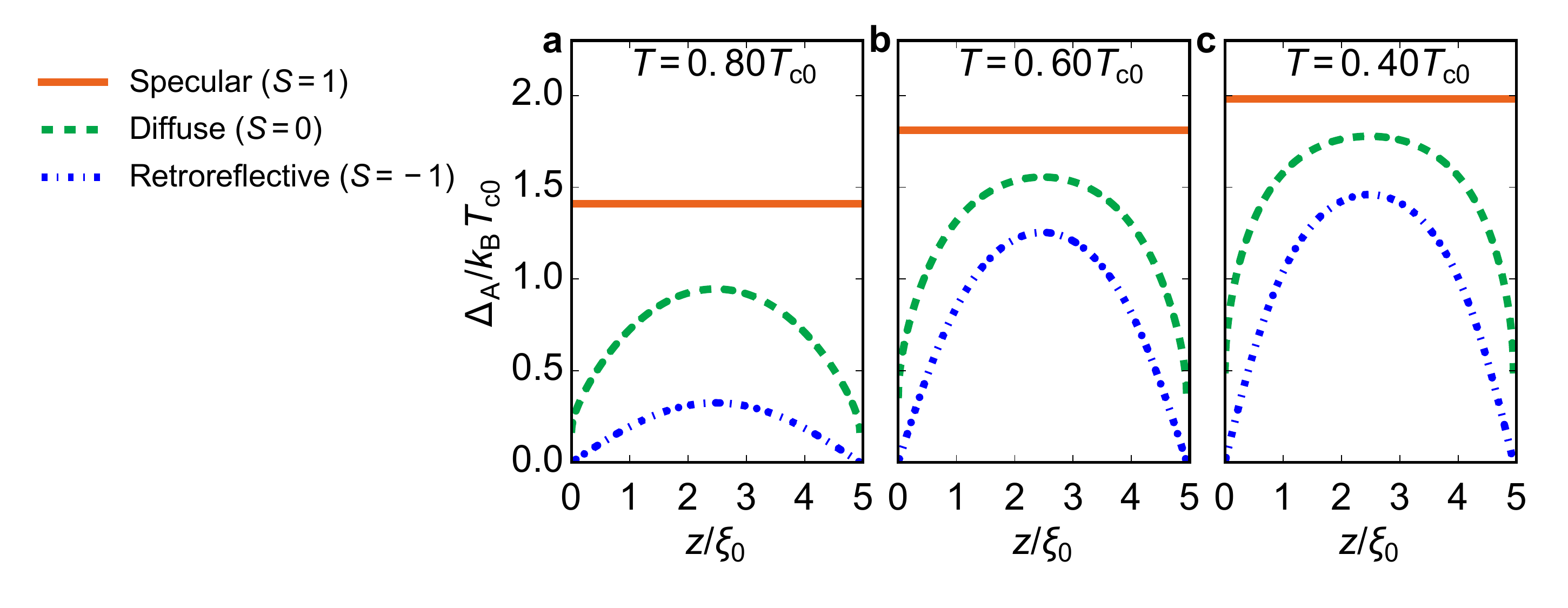}
	\caption{\label{fig:gap_prof} \textbf{Spatial dependence of energy gap in slab-shaped cavity.} With specular boundary condition for quasiparticles, the energy gap $\Delta_{\mathrm{A}}$ remains constant equalling the corresponding bulk value. With diffuse boundary condition, the gap is suppressed at the walls resulting in spatial dependence along the $z$-axis, whereas maximally pair-breaking retroreflective boundary condition fully suppresses the gap at the walls. The values shown here at three different temperatures (\textbf{a, b, c}), where $T_{\mathrm{c0}}$ is the bulk superfluid transition temperature, correspond to effective cavity height $D/\xi_0=4.95$ ($P=5.50\,$bar in our cavity) and are based on quasiclassical weak-coupling theory for superfluid $^3$He~\cite{QuasiClass}, following the calculational methods presented in the text.}
\end{figure*}

\begin{figure*}
	\includegraphics[width=\textwidth]{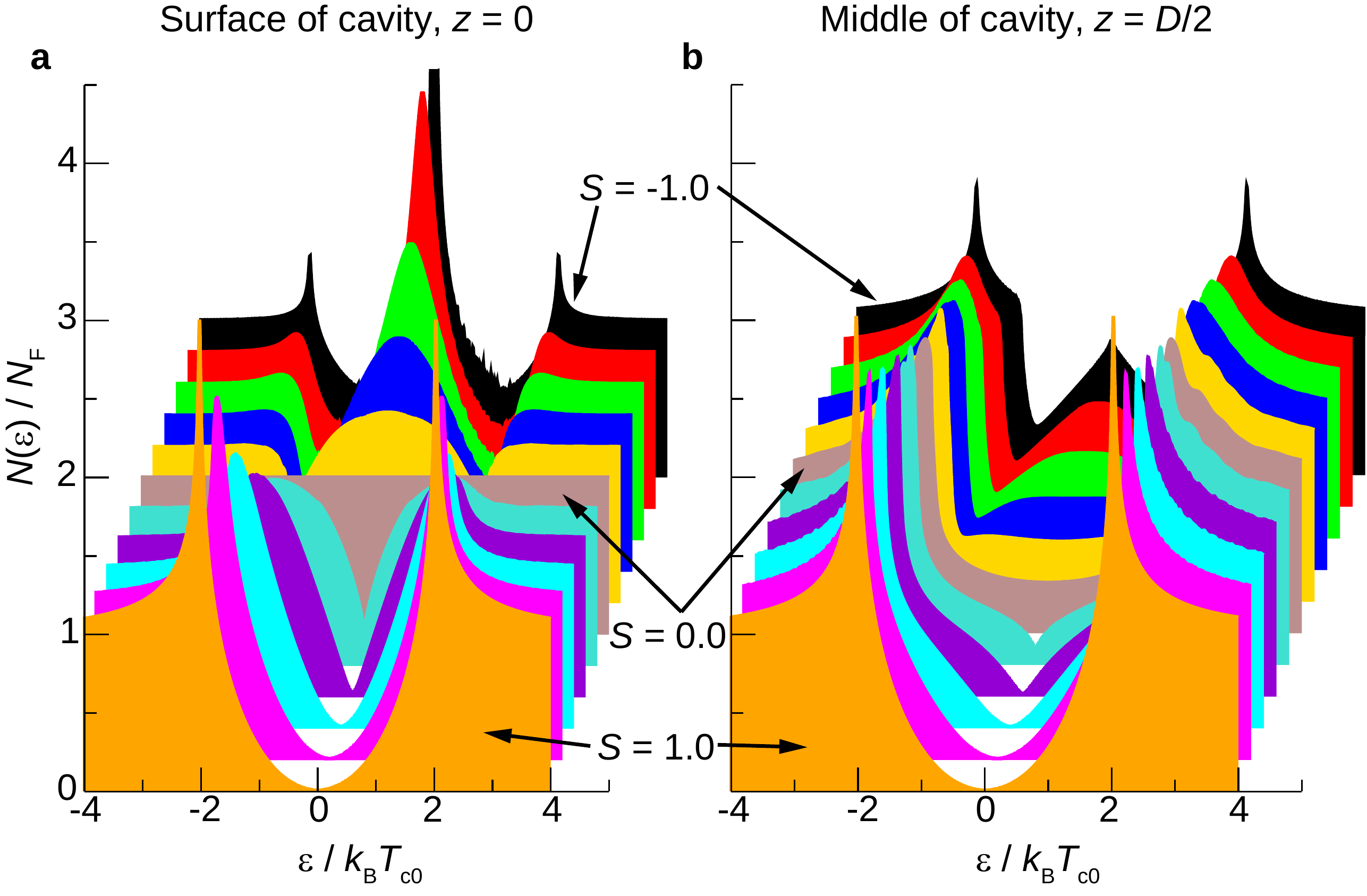}
	\caption{\label{fig:DOS_vs_S} \textbf{Density of states of $^3$He-A at the surface and in the middle of the cavity.} Just like the energy gap, the density of states $N(\epsilon)$, where $\epsilon$ is the quasiparticle energy, evolves as a function of the surface specularity. Specular boundary condition results in the bulk DOS both at the surface (\textbf{a}) and in the middle of the cavity (\textbf{b}). In bulk, the finite density of sub-gap states is a consequence of nodes in the gap structure of $^3$He-A in momentum space~\cite{VW}. As the specularity decreases and the energy gap gradually suppresses towards the surface (see Supplementary Fig.~\ref{fig:gap_prof}), the density of sub-gap states increases throughout the cavity. At the surface, the fully diffuse scattering gives flat band of DOS whereas the maximally pair-breaking retroreflective scattering results in significant accumulation of zero-energy bound states. The values shown here are scaled with the normal state DOS, $N_{\mathrm{F}}$, and averaged over the full Fermi surface. In calculations we use cavity height $D/\xi_0 = 5.00$ and temperature $T = 0.20T_{\mathrm{c0}}$. The step size in specularity is 0.2 between consequent colours.}
\end{figure*}

The boundary condition that connects the propagators of the incoming $(\mathbf{p}_{\mathrm{in}})$ and outgoing $(\mathbf{p}_{\mathrm{out}})$ trajectories at the surface is written as~\cite{Nagai_JPSJ2008}
\begin{equation}
	\widehat{A} \widehat{g} (\mathbf{p}_{\mathrm{out}}) = \widehat{g}(\mathbf{p}_{\mathrm{in}}) \widehat{A},\qquad
	\widehat{A} = \frac{1+i\widehat{\sigma}}{1-i\widehat{\sigma}}.
\end{equation}
The surface self-energy $\widehat{\sigma}$ is momentum-independent and computed self-consistently from 
\begin{equation}
	\widehat{\sigma} = \frac{2W}{1 + \widehat{\sigma}^2} \left( \widehat{g}_{\mathrm{surf}} - \widehat{\sigma} \right).
\end{equation}
Here the surface propagator for specular (or retro) reflection is averaged over the momenta parallel to surface, which we can write as integration over incoming or outgoing momenta on the Fermi surface (FS)
\begin{equation}
	\widehat{g}_{\mathrm{surf}} = \langle \widehat{g}_{\mathrm{in}} \rangle_{||} 
	= \langle \widehat{g}_{\mathrm{out}} \rangle_{||}
	= \langle |\hat{v}_z| \widehat{g}_{\mathrm{in}} \rangle^{}_{\mathrm{FS-}}
	= \langle |\hat{v}_z| \widehat{g}_{\mathrm{out}} \rangle^{}_{\mathrm{FS+}}.
\end{equation}
The integral is normalised with the area of FS in the reflective plain, $\pi p_{\mathrm{F}}^2$, i.e.,
\begin{equation}
	\langle \dots \rangle_{||} = \frac{1}{\pi p_{\mathrm{F}}^2} \iint dp_x dp_y \dots = \frac{1}{\pi}
	\int\limits_{\hat{v}_z > 0\,\textrm{or}\,\hat{v}_z < 0} d\Omega_{\mathrm{F}} |\hat{v}_z| \dots 
\end{equation}
In the normal state $\widehat{g}_{\mathrm{surf}} = \widehat{g}^{}_{\mathrm{N}} = -i \; \mathrm{sgn}(\epsilon_n) \widehat{\tau}_3$, and solution for surface self-energy is $\widehat{\sigma} = \sigma \widehat{g}^{}_{\mathrm{N}}$ with $(1+\sigma)\sigma = 2W$.

In the presented calculations we assume that the quality of both surfaces of the cavity is approximately equal and thus characterised by the same specularity parameter $S$, as is true with silicon-silicon sample containers having even $^4$He preplating between the surfaces. Example energy gap profiles corresponding to different specularities are shown in Supplementary Fig.~\ref{fig:gap_prof} and density of states calculations in Supplementary Fig.~\ref{fig:DOS_vs_S}.

\clearpage 

\section{Strong-coupling corrections}
\label{sec:strong-coupling}

All calculations of energy gap $\Delta_{\mathrm{A}}$ in the main text are based on quasiclassical weak-coupling theory adjusted for strong-coupling effects near $T_{\mathrm{c0}}$. This is straightforward since the energy gap in the G-L regime is written as~\cite{VW}
\begin{equation}
	\label{eq:gap_G-L}
	\Delta_{\mathrm{A}}^2 = \frac{\Delta C_{\mathrm{A}}}{C_{\mathrm{N}}} \left(\pi k_{\mathrm{B}} T_{\mathrm{c0}}\right)^2 \left(1-T/T_{\mathrm{c0}}\right),
\end{equation}
where $\Delta C_{\mathrm{A}}$ refers to change in the specific heat at superfluid transition and $C_{\mathrm{N}}$ is the normal state specific heat. From this the initial slope of the gap acquires a simple form (see also Supplementary Eq.~(\ref{eq:IS_Delta})):
\begin{equation}
	\label{eq:IS_G-L}
	\mathrm{IS}_{\Delta}^{\mathrm{bulk}} = \frac{\Delta C_{\mathrm{A}}}{C_{\mathrm{N}}} \left(\pi k_{\mathrm{B}} T_{\mathrm{c0}}\right)^2.
\end{equation}
Thus, the ratio between the reported pressure-dependent value of the specific-heat jump $\frac{\Delta C_{\mathrm{A}}}{C_{\mathrm{N}}}$~\cite{StrongCoup_2007} and its weak-coupling value 1.188 can be used as a pressure-dependent correction factor for both weak-coupling $\Delta_{\mathrm{A}}^2$ and $\mathrm{IS}_{\Delta}^{\mathrm{bulk}}$. We assume that Supplementary Eq.~(\ref{eq:gap_G-L}) is also valid for non-bulk values of the gap, $\left<\Delta_{\mathrm{A}}^2(z)\right>$, with suppressed superfluid transition temperature $T_{\mathrm{c}}$ to make a similar trivial correction for non-specular boundary conditions, see Fig.~2c in the main text.

However, the strong-coupling effects in general depend both on temperature as well as on pressure. For $^3$He-B these effects have been included in the weak-coupling-plus model~\cite{QuasiClass}. The corresponding calculations~\cite{Sauls_private} extended for thermodynamic properties of bulk $^3$He-A (such as the maximum energy gap $\Delta_{\mathrm{A}}$) show very good agreement with our measurements under ``specular'' boundary conditions, see Supplementary Fig.~\ref{fig:strong-coupling}. These calculations assume the gap to be uniform, i.e., do not take into account the nodes in the gap structure of $^3$He-A, but it is to be expected that the inclusion of the nodal excitations will further improve the agreement. Similar calculations for different surface scattering boundary conditions do not yet exist, but it is our assessment that the low-temperature deviation between our ``diffuse'' measurement and the trivially corrected weak-coupling $S=0.10$ calculation visible at 5.50\,bar in Fig.~2c in the main text evidences the need for the same kind of temperature-dependent strong-coupling corrections when the average gap is suppressed by confinement.

\begin{figure}
	\centering
	\includegraphics[width=0.8\textwidth]{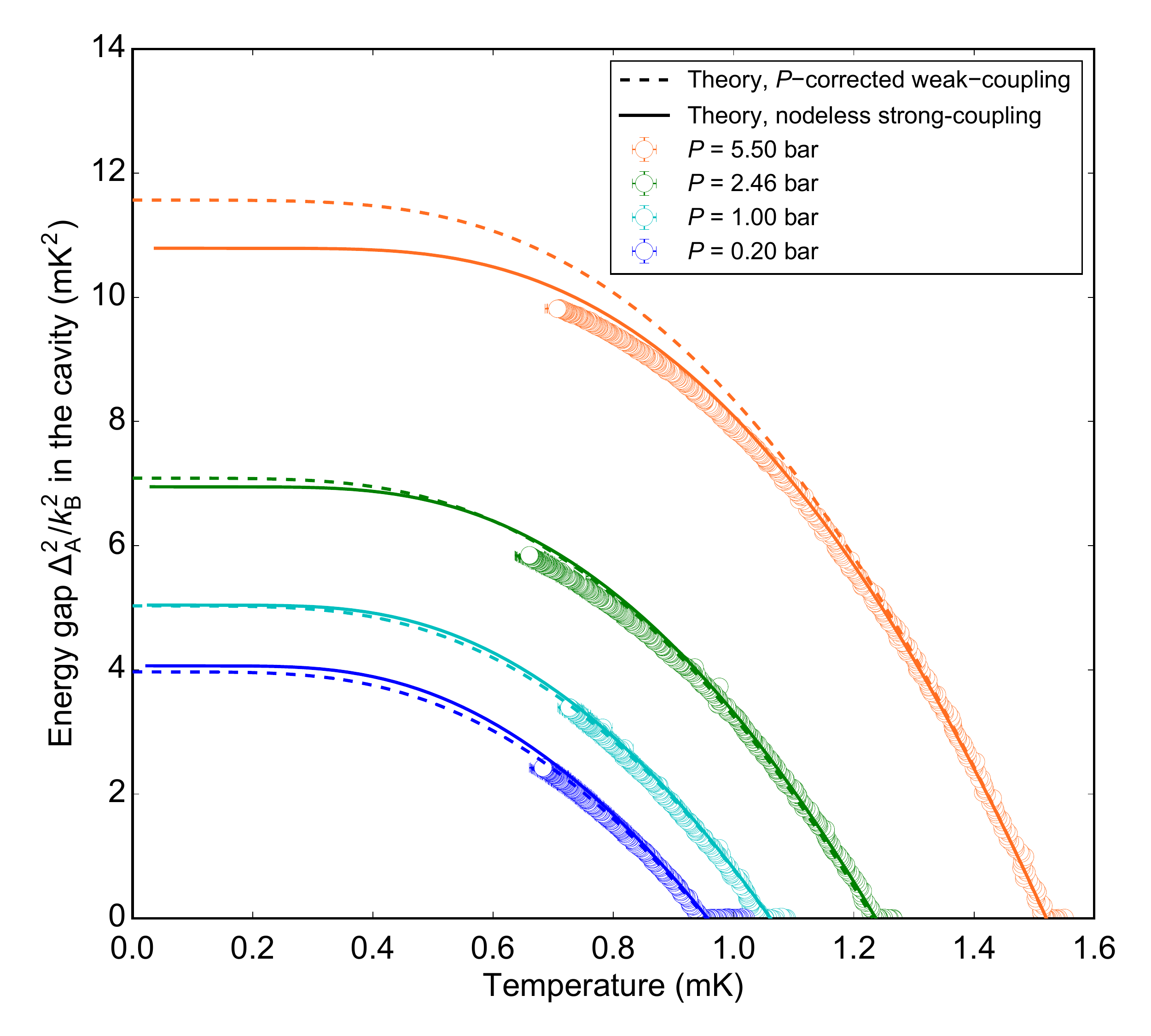}
	\caption{\label{fig:strong-coupling} \textbf{Strong-coupling corrections to bulk energy gap.} Comparison of measured energy gap in the cavity under ``specular'' boundary condition shows very good agreement with bulk energy gap $\Delta_{\mathrm{A}}$ including temperature-dependent strong-coupling corrections (solid lines), see text. The result neglecting the temperature dependence of strong-coupling corrections shown in Fig.~2c in the main text is also shown here for comparison (dashed lines).}
\end{figure}

\clearpage

\section{Determination of $T_{\mathrm{c}}$ in presence of solid $^3$He surface boundary layer}
\label{sec:solid3He}

\subsection{Sample magnetisation}
\label{sec:magnetisation}

With pure $^3$He in the sample container, localised $^3$He atoms next to a wall form a magnetic surface boundary layer~\cite{Ahonen_1976,2D_magnets}. The susceptibility of this layer has been shown to obey Curie-Weiss law $\chi_{\mathrm{s}} = C/(T-\theta)$, where $C$ is the Curie constant, at high temperatures $T \gtrsim 1\,$mK with positive Weiss temperature $\theta$, characteristic for systems with a ferromagnetic tendency~\cite{Freeman_1990}. The rest of the cavity is filled by liquid having a normal state $^3$He susceptibility $\chi^{}_{\mathrm{N}}$, whether in normal or in superfluid $^3$He-A state. Total magnetisation of the sample is written as $M = M_{\mathrm{s}} + M_{\mathrm{FL}}$, where $M_{\mathrm{s}} \propto \chi_{\mathrm{s}}$ represents the magnetisation of the solid layer and $M_{\mathrm{FL}} \propto \chi^{}_{\mathrm{N}}$ is the liquid's magnetisation. The Fermi liquid susceptibility is well described by the phenomenological expression given by Dyugaev, $\chi^{}_{\mathrm{N}} \propto 1/\sqrt{T^2 + T_{\mathrm{F}}^{**2}}$~\cite{Dyugaev_sovsci}, where effective Fermi temperature $T^{**}_{\mathrm{F}}$ is a density-dependent parameter~\cite{2D_Fermi_liquid,2D_Fermi_liquid_RHUL,susc_3He}. In our experimental temperature range $T\ll T^{**}_{\mathrm{F}} \approx 200-300\,$mK, the liquid magnetisation is constant, thus reducing the temperature dependence of $M$ to result purely from the solid layer.

Temperature dependence of total magnetisation $M$ in the cavity at two pressures is shown in Supplementary Fig.~\ref{fig:CW}a. Magnetisation is determined from Lorentzian fits to Fourier-transformed data. Unlike in Refs.~\cite{2D_solid_3He,2D_solid_3He_RHUL}, we do not observe line broadening as a function of temperature possibly due to extreme smoothness of the silicon surfaces. The constant magnetisation of the liquid is measured independently when having 32\,\textmu mol/m$^2$ of $^4$He in the sample, i.e., ``diffuse'' boundary condition and no evidence of temperature-dependent magnetisation since $^4$He atoms have replaced all the localised $^3$He on the walls.

Plotting the inverse of solid magnetisation against temperature, as shown in Supplementary Fig.~\ref{fig:CW}b, lets us determine the Weiss temperature as the intersection between the linear fit and $1/M_{\mathrm{s}}=0$. This gives $\theta$ between 0.65 and 0.75\,mK, which is consistent with earlier reported values changing between 0.3 and 0.8\,mK~\cite{Ahonen_1976,Freeman_1990,2D_solid_3He,solid3He_fluorocarbon}. Possible systematic error follows from the shape of frequency spectra at $T\gtrsim 0.9T_{\mathrm{c0}}$ where the spectral closeness of the signals arising from the bulk markers distorts the shape of the cavity signal, thus increasing the uncertainties in Lorentzian fitting.

We use the ferromagnetic high-temperature series expansion (HTSE) for triangular lattice up to 9th order in $J_{\chi}/T$ to fit the low-temperature values of $M_{\mathrm{s}}$ to be used in further analysis, Supplementary Fig.~\ref{fig:CW}a. The exchange coefficient is defined as $J_{\chi}=\theta /3$. We have used the series expansion coefficients corresponding to Heisenberg model taking into account only two-particle exchange as given in Ref.~\cite{HTSE_1967}. These agree with the coefficients found for cyclic multiple-spin-exchange (MSE) model when only the dominant two and three-particle exchange processes (the effective Heisenberg exchange) are considered~\cite{HTSE_Roger}. HTSE has earlier been used to successfully model solid $^3$He layer on graphite~\cite{2D_solid_3He_RHUL,2D_Heis_Ferr,2D_solid_3He_thermod,2D_magnets}. To avoid problems due to uncertain values of $M$ near and above the bulk superfluid transition temperature $T_{\mathrm{c0}}$, only the values extracted at temperatures $T\lesssim 0.9T_{\mathrm{c0}}$ are used in fitting. This way we find $J_{\chi}\approx 0.15\,$mK which corresponds to the reported values of $\theta$.

\begin{figure*}
	\includegraphics[width=\textwidth]{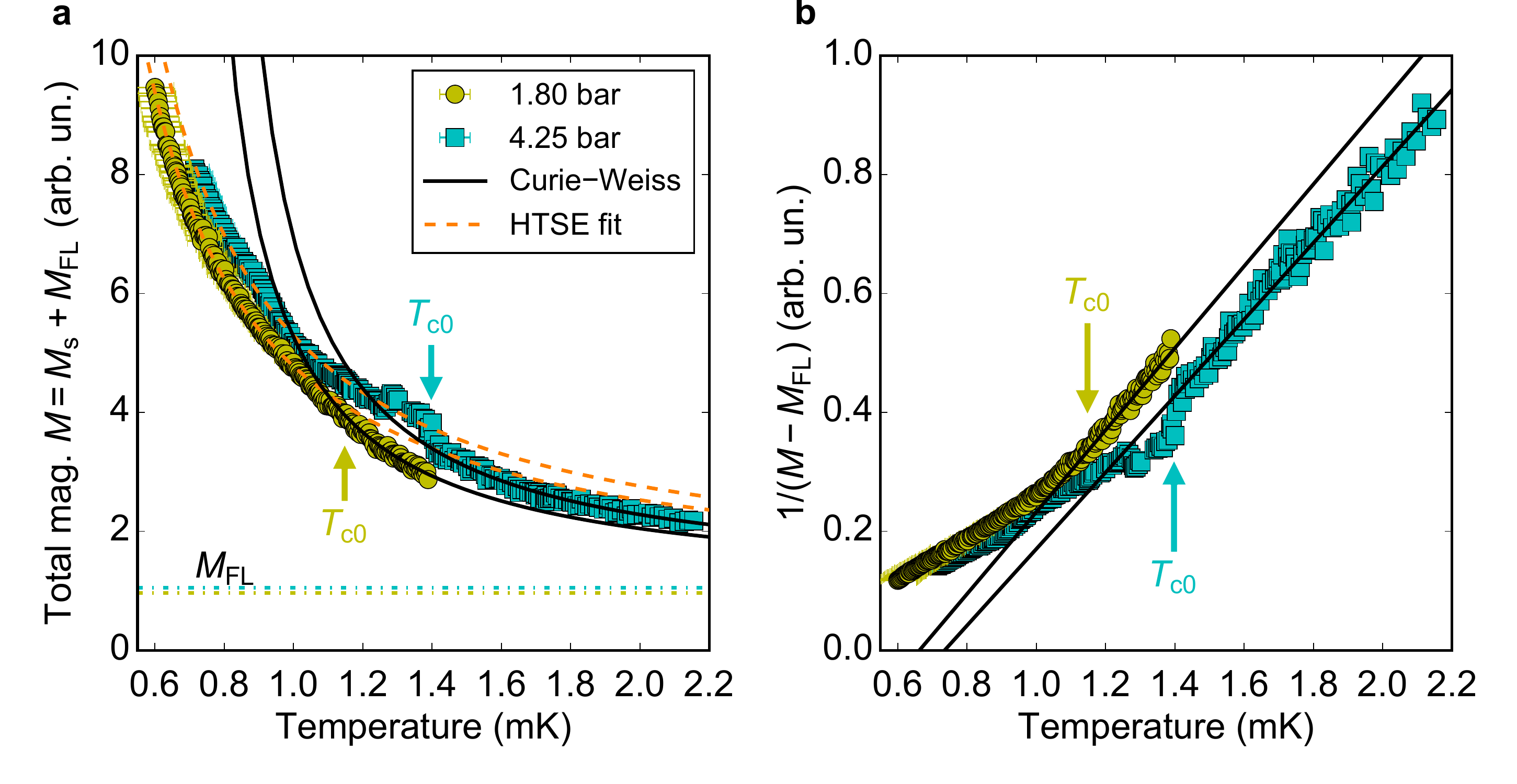}
	\caption{\label{fig:CW} \textbf{Magnetisation with solid layer of $^3$He on the cavity walls.} \textbf{a,} Total magnetisation $M$ versus temperature at two different pressures. Independently determined liquid magnetisations $M_{\mathrm{FL}}$ are shown as horizontal dash-dotted lines and bulk superfluid transition temperatures are marked with arrows. At temperatures $T\gtrsim 1\,$mK the magnetisations are seen to follow the Curie-Weiss law and at temperatures $T\lesssim 0.9T_{\mathrm{c0}}$ the data are fit using HTSE (the deviations from these fits at higher temperatures are due to systematic errors in determination of $M$). Small pressure dependence of $M$ on top of the almost constant $M_{\mathrm{FL}}$ results from the compression of the solid layer with increasing pressure~\cite{2D_Fermi_liquid,solid3He_fluorocarbon,solid_mag_aerogel}. \textbf{b,} The Weiss temperature $\theta$ is defined as the abscissa of the linear fit (solid lines) to the $1/M_{\mathrm{s}}$ versus temperature at $1/M_{\mathrm{s}} = 0$. The visible kink just below $T_{\mathrm{c0}}$, especially seen in 4.25\,bar data, results from Lorentzian fitting of three close peaks (cavity signal and two bulk marker signals) being unreliable in determining the magnetisations of the individual features.}
\end{figure*}

\clearpage

\subsection{Extraction of superfluid transition temperature}
\label{sec:Tc_solid3He}

Due to the two-dimensional nature of the magnetic $^3$He surface boundary layer, the local field of the oriented neighbouring spins results in negative frequency shift $\Delta f_{\mathrm{s}} \propto M_{\mathrm{s}}$ in NMR measurements when $\mathbf{H}_0 = H_0\hat{\mathbf{z}}$ is oriented normal to the surface~\cite{Freeman_1990,2D_solid_3He}. The atomic exchange between the solid and liquid components results in a single peak in the spectrum. The amplitude, frequency, and line shape of this composite signal depend on various factors, such as the intrinsic magnetisations and relaxation rates of the solid and the liquid as well as their exchange rates and the relative frequency shift~\cite{Freeman_1990}. The detected frequency shift of the composite peak, $\Delta f$, is determined as a weighted average of the intrinsic frequency shifts $\Delta f_{\mathrm{s}}$ and $\Delta f_{\mathrm{sf}}$ (superfluid)~\cite{Freeman_1988}:
\begin{equation}
	\label{eq:Df_solid}
	\Delta f = \frac{M_{\mathrm{s}}}{M}\Delta f_{\mathrm{s}} + \frac{M_{\mathrm{FL}}}{M}\Delta f_{\mathrm{sf}}.
\end{equation}
In normal state we have $\Delta f_{\mathrm{sf}} = 0$ and in the superfluid state its sign and magnitude depend on the superfluid phase, whereas the solid-induced temperature-dependent shift $\Delta f_{\mathrm{s}}$ is present regardless of the liquid being normal or superfluid. In the nanofluidic cavity of our sample cell, the ratio of solid-to-liquid magnetisation is high, $M_{\mathrm{s}} > M_{\mathrm{FL}}$. Thus, the solid frequency shift dominates the overall temperature dependence, masking the straightforward detection of superfluid transition in the cavity. Isolation of the signal arising from the solid is therefore a necessary prerequisite for the extraction of $T_{\mathrm{c}}$. See Supplementary Fig.~\ref{fig:fshift_comparison} for measured frequency shifts corresponding to three different scattering boundary conditions.

We adopt two methods, with consistent results, to extract the superfluid $T_{\mathrm{c}}$ in the cavity in the presence of high background frequency shift arising from the solid $^3$He:

\noindent (1) Direct comparison of frequency shifts measured at different pressures. At zero pressure the superfluidity is completely suppressed down to the lowest temperatures investigated. Thus, the zero-pressure frequency shift is well-described by HTSE fit over the full temperature range, and the superfluid transition at any higher pressure is identified as the temperature where the frequency shift deviates from this zero-pressure fit. This method is presented in Supplementary Fig.~\ref{fig:solid3He_Tc}a,b and relies on the assumption that the solid magnetisation, and thus the resulting frequency shift, is not dependent on pressure.

\noindent (2) Extraction of superfluid frequency shift by removing the solid effect. First, we write the solid frequency shift as $\Delta f_{\mathrm{s}} = C_{\mathrm{s}}M_{\mathrm{s}}$, where $C_{\mathrm{s}}$ is a proportionality constant. Now we get a two-domain frequency shift by using Supplementary Eq.~(\ref{eq:Df_solid}):
\begin{eqnarray}
	\label{eq:solid_sf_fshift}
	\Delta f &=& C_{\mathrm{s}}\frac{\left(M - M_{\mathrm{FL}}\right)^2}{M}, \textrm{ when } T > T_{\mathrm{c}}, \\
	\Delta f &=& C_{\mathrm{s}}\frac{\left(M - M_{\mathrm{FL}}\right)^2}{M} + \frac{M_{\mathrm{FL}}}{M}\Delta f_{\mathrm{sf}}, \textrm{ when } T < T_{\mathrm{c}}.
\end{eqnarray}
Thus, the frequency shift in the superfluid alone is
\begin{equation}
	\label{eq:Df_sf}
	\Delta f_{\mathrm{sf}} = \frac{M}{M_{\mathrm{FL}}}\left( \Delta f - C_{\mathrm{s}}\frac{\left(M - M_{\mathrm{FL}}\right)^2}{M} \right).
\end{equation}
We determine $C_{\mathrm{s}}$ from a linear fit to the the measured $\Delta f$ against $\left(M - M_{\mathrm{FL}}\right)^2/M$ in range $T \lesssim 0.9T_{\mathrm{c0}}$ to as low temperature as the data show linearity (at zero pressure to the lowest temperature investigated). This can be done individually for each data set (Supplementary Fig.~\ref{fig:solid3He_Tc}c), or, if assuming pressure-independence, we can use the zero-bar $C_{\mathrm{s}}$ at every pressure (Supplementary Fig.~\ref{fig:solid3He_Tc}d).

Both methods (1) and (2) extract a clear break in the temperature dependence of the frequency shift, making the determination of superfluid transition straightforward. The values of $T_{\mathrm{c}}$ corresponding to solid $^3$He boundary condition plotted in Fig.~3 in the main text are based on method (2), using individually determined $C_{\mathrm{s}}$. However, whichever method we use, the values coincide with each other. Due to uncertainties in determination of $M_{\mathrm{s}}$ and $C_{\mathrm{s}}$, we do not consider the extracted temperature dependence of the superfluid frequency shift (Supplementary Fig.~\ref{fig:solid3He_Tc}c,d) to be reliable. A significant improvement is possible by conducting measurements in lower magnetic fields, which both increases the absolute value of $\Delta f_{\mathrm{sf}}$ and reduces the absolute magnitude of the dipolar frequency shift $\Delta f_{\mathrm{s}}$ arising from the solid.

\begin{figure*}
	\includegraphics[width=\textwidth]{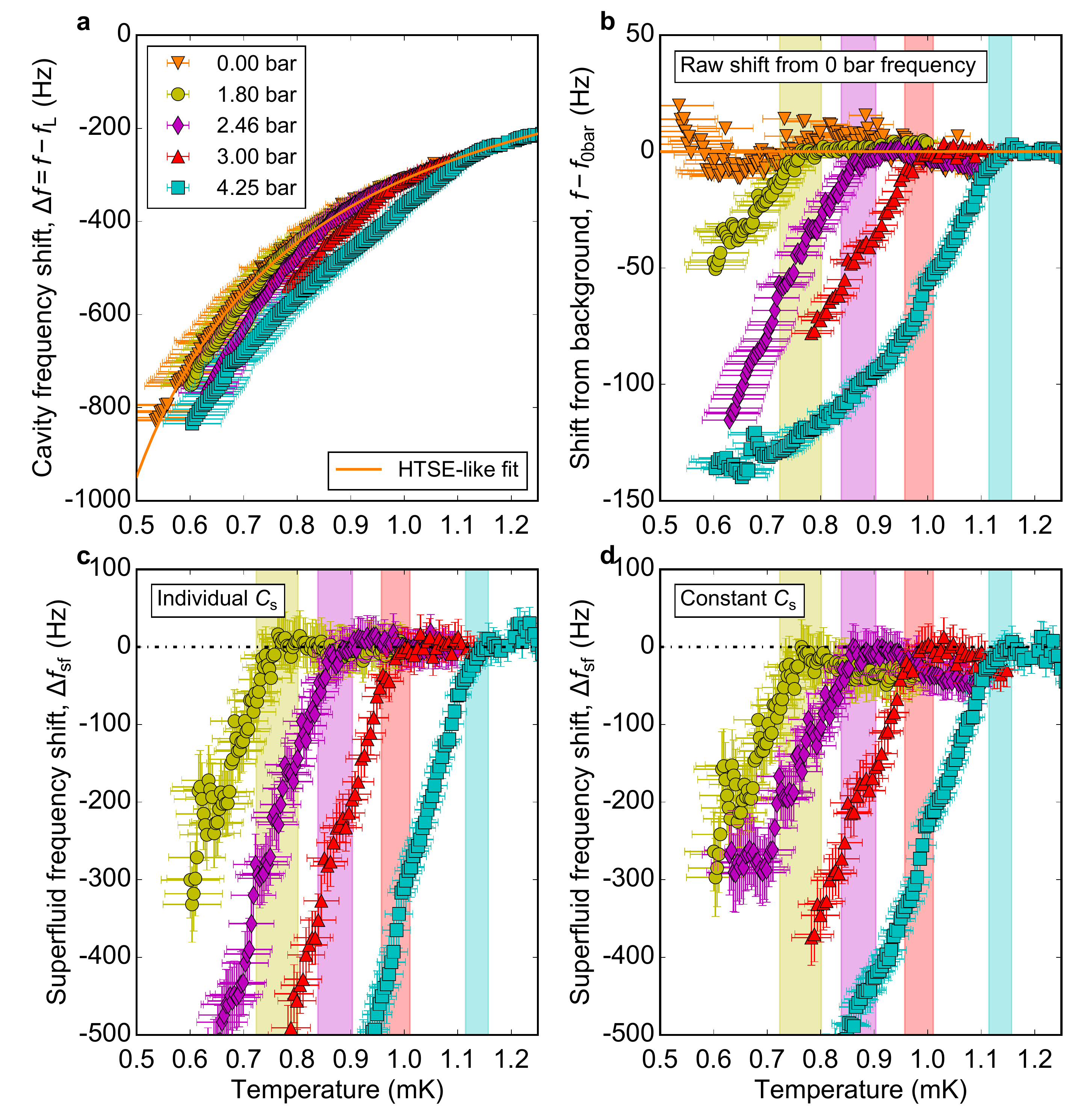}
	\caption{\label{fig:solid3He_Tc} \textbf{Superfluid frequency shift in the cavity with solid $^3$He on the walls.} \textbf{a,} NMR frequency shifts of the cavity signal at higher pressures deviate from the low-pressure values at certain temperatures indicating a superfluid transition in the liquid (method (1)). Solid line is a fit to 0\,bar data using high-temperature series expansion as a model for the solid magnetisation. To compensate the small pressure dependence in $M_{\mathrm{s}}$, the Larmor frequency $f_{\mathrm{L}}\approx 967\,$kHz of each dataset is adjusted by less than 20\,Hz to make $\Delta f$ agree with the 0\,bar HTSE fit at $0.9T_{\mathrm{c0}}$ of each pressure. \textbf{b,} The temperatures at which the deviations occur are clearly detected when the ``background'' frequency shift at 0\,bar is subtracted from the data to determine $T_{\mathrm{c}}$. \textbf{c--d,} The pure superfluid frequency shifts without the solid contribution can be extracted using method (2) with individually determined $C_{\mathrm{s}}$ or with constant zero-bar $C_{\mathrm{s}}$ as described in the text. The latter way clearly gives smaller $\Delta f_{\mathrm{sf}}$. The coloured vertical bands in \textbf{b}, \textbf{c}, and \textbf{d} indicate the values of $T_{\mathrm{c}}$, with uncertainty, determined using the method illustrated in \textbf{c}. It is seen that all the techniques showcased here give a consistent result.}
\end{figure*}

\clearpage

\section{Pair breaking at the surface}
\label{sec:magnetic_scattering}

\subsection{Momentum scattering}
\label{sec:momentum_scattering}

The scattering of quasiparticles from surface and the consequential pair breaking are incorporated into the quasiclassical theory in a form of boundary conditions for the propagator (see also Supplementary Note.~\ref{sec:theory_OP}). An efficient way to write them is to connect the coherence amplitude $\Gamma(\mathbf{p}_{\mathrm{out}}) i\sigma_y$ on the outgoing trajectory with the incoming amplitudes $\gamma(\mathbf{p}_{\mathrm{in}}) i\sigma_y$ through scattering matrix~\cite{Eschrig_2009}, 
\begin{equation}
	\Gamma(\mathbf{p}_{\mathrm{out}}) i\sigma_y = \mathcal{S} \, \left[ \gamma(\mathbf{p}_{\mathrm{in}}) i\sigma_y \right] \, \tilde{\mathcal{S}},
\end{equation}
where $\mathcal{S}$ is the normal state scattering matrix in the particle sector and $\tilde{\mathcal{S}}$ in the hole sector. Coherence amplitudes carry the information about the structure of the order parameter, and the difference between the asymptotic value $\gamma(\mathbf{p}_{\mathrm{out}})$ and the initial scattered value $\Gamma(\mathbf{p}_{\mathrm{out}})$ of coherence amplitude on the outgoing trajectory can give an indication of the pair-breaking properties of the surface. Far from the surface, the asymptotic value on any trajectory $\mathbf{k}$ is
\begin{equation}
	\gamma(\mathbf{k}) \equiv \gamma_0 + \bm{\gamma}_{\mathrm{t}}(\mathbf{k}) \cdot \bm{\sigma} = \gamma_0 + \gamma_x \sigma_x +\gamma_y \sigma_y + \gamma_z \sigma_z.
\end{equation}
Here $\gamma_0$ and $\bm{\gamma}_{\mathrm{t}}(\mathbf{k}) = \gamma_x(\mathbf{k}) \hat{\mathbf{x}} + \gamma_y(\mathbf{k}) \hat{\mathbf{y}} + \gamma_z(\mathbf{k}) \hat{\mathbf{z}}$ are the singlet and triplet components in spin space, respectively, and $\bm{\sigma} = \sigma_x \hat{\mathbf{x}} + \sigma_y \hat{\mathbf{y}} + \sigma_z \hat{\mathbf{z}}$ is the Pauli vector consisting of Pauli matrices $\sigma_x$, $\sigma_y$, and $\sigma_z$. For unitary phases coherence amplitudes reflect symmetry of order parameter $\Delta (\mathbf{k})$~\cite{Vorontsov_PhilA}: $\gamma_0 i\sigma_y \propto \Delta_0 i\sigma_y$ and $\left( \bm{\gamma}_{\mathrm{t}}(\mathbf{k}) \cdot \bm{\sigma} \right) i\sigma_y \propto \Delta_{\mathrm{t}} (\mathbf{k}) = \left( \mathbf{d}(\mathbf{k}) \cdot \bm{\sigma} \right) i\sigma_y$, where $\mathbf{d}(\mathbf{k})$ is the order-parameter vector.

Now we consider $^3$He-A (spin-triplet pairing) with $\mathbf{d}(\mathbf{k}) \perp \mathbf{H}_0 = H_0 \hat{\mathbf{z}}$ relevant to our experiments~\cite{VW}:
\begin{equation}
	\mathbf{d}(\mathbf{k}) = \hat{\mathbf{x}} \Delta_{\mathrm{A}} (\hat{k}_x + i \hat{k}_y) \quad \propto \quad \vert \uparrow\uparrow \rangle + \vert \downarrow\downarrow \rangle.
\end{equation}
For the purely momentum scattering one can employ the surface roughness averaging over incoming directions that leads to 
\begin{equation}
	\label{eq:momentum_scatt}
	\Gamma(\mathbf{p}_{\mathrm{out}}) = |S| \gamma(\mathbf{p}_{\mathrm{in}}) + (1-|S|) \langle \gamma \rangle_\parallel = |S| \gamma(\mathbf{p}_{\mathrm{in}}),
\end{equation}
where the coherence parameter $0 \leq |S| \leq 1$, as defined below Supplementary Eq.~(\ref{eq:specularity}), describes the relative amount of quasiparticles scattering coherently (non-diffusively) between $\mathbf{p}_{\mathrm{in}}$ and $\mathbf{p}_{\mathrm{out}}$, and for A phase the momentum average over all in-plane directions (diffuse scattering) gives zero, $\langle \gamma \rangle_\parallel \propto \langle p_x + ip_y \rangle_\parallel = 0$.

The scattering is associated with a relative phase change $\phi$ on the order parameter ``seen'' by the quasiparticle on the incoming and the outgoing trajectories: for skew scattering in the A phase $\Delta (\mathbf{p}_{\mathrm{out}}) = e^{-i\phi}\Delta (\mathbf{p}_{\mathrm{in}})$. As shown in Ref.~\cite{Vorontsov_PhilA}, this scattering phase results in the energy spectrum of surface-bound states: $E_{\mathrm{SBS}} / \Delta_{\mathrm{A}} = \pm \cos \phi / 2$. For purely momentum scattering in $^3$He-A, $\phi$ corresponds to $x$-$y$ plane rotation in momentum space. Fully specular scattering ($|S| = 1, \phi = 0$) generates no pair breaking and no additional sub-gap energy states at the surface, giving the relation $\Gamma(\mathbf{p}_{\mathrm{out}}) = \gamma (\mathbf{p}_{\mathrm{in}}) = \gamma (\mathbf{p}_{\mathrm{out}})$ for the coherence amplitudes. The completely backscattering (retroreflective) surface $\mathbf{p}_{\mathrm{out}} = -\mathbf{p}_{\mathrm{in}}$ with $|S| = 1$ and $\phi = \pi$ leads to largest weight of zero-energy surface states, $E_{\mathrm{SBS}} = 0$, and gives $\Delta (\mathbf{p}_{\mathrm{out}}) = - \Delta (\mathbf{p}_{\mathrm{in}})$ on all scattering trajectories, fulfilling the condition of coherence amplitudes generating maximal pair breaking and suppression of $T_{\mathrm{c}}$: $\Gamma(\mathbf{p}_{\mathrm{out}}) = \gamma (\mathbf{p}_{\mathrm{in}}) = - \gamma (\mathbf{p}_{\mathrm{out}})$. In purely diffuse case ($|S|=0$) the scattering angle $\phi$ is fully diffuse, i.e., evenly spread over the entire phase circle between $\phi = 0$ and $\phi = 2\pi$, resulting in all the possible bound state energies filling up the gap uniformly. Partially diffuse scattering has $0 < |S| < 1$. All this is visible in the calculations of DOS, Supplementary Fig.~\ref{fig:DOS_vs_S}.

In general, the quasiparticle reflection picks up relative phase $\phi$, responsible for bound-state energies, from both momentum and magnetic degrees of freedom between the incoming and outgoing trajectories. We investigate below whether the spin dependence in scattering can increase the density of surface-bound states close to zero energy ($\phi = \pi$) and thus result in more than diffuse pair breaking even in the absence of partial retroreflection by the surface. For the A phase with scattering path $\gamma(\mathbf{p}_{\mathrm{out}}) = \gamma (\mathbf{p}_{\mathrm{in}})$, maximal pair breaking will occur when $\Gamma (\mathbf{p}_{\mathrm{out}}) = -1\cdot\gamma (\mathbf{p}_{\mathrm{in}})$.

\subsection{Magnetic scattering from polarized surface}
\label{sec:polarized_surface}

The magnetically active highly-polarizable solid layer of $^3$He on the surface can affect the spin structure of the superfluid phase stabilised in the cavity. These effects will originate from the spin-dependent scattering of the quasiparticles from the solid layer and should be included in the boundary condition for the quasiclassical propagator. On the general grounds one expects that such effects could be magnetically anisotropic and strongly depend on the spin structure of the superfluid phase and the orientation of the magnetic field. Here we are particularly interested in the role of magnetic scattering on the suppression of $T_{\mathrm{c}}$.

Magnetic scattering in this formalism is included through the spin-dependent part $\mathcal{M}$ of the $\mathcal{S}$-matrix. This matrix has the form~\cite{Millis_1988} 
\begin{equation}
	\mathcal{M} = 
	\begin{pmatrix} 
		e^{-i \delta_{\uparrow \mathbf{m}}} & \\ 
		& e^{-i\delta_{\downarrow \mathbf{m}}} 
	\end{pmatrix} 
	= e^{-i\vartheta_0} e^{-i( \hat{\mathbf{m}} \cdot \bm{\sigma} ) \vartheta / 2}
\end{equation}
and it describes the phase difference $\vartheta = \delta_{\uparrow \mathbf{m}} -\delta_{\downarrow \mathbf{m}}$ that spins up and down (in $\hat{\mathbf{m}}$-basis) acquire when they scatter off the (classically) magnetically-polarized surface. Here unit vector $\hat{\mathbf{m}}$ refers to the polarization axis of magnetisation of the surface layer. The hole sector matrix is obtained assuming particle-hole symmetry $\tilde{\mathcal{M}} = \mathcal{M}^*$. The spin-dependent phase difference can be related, for example, to the exchange coupling in the ferromagnetic layer~\cite{Tokuyasu_1988}, or one can consider a possibility of Kondo-like resonant scattering that may enhance the effective exchange interaction and produce large phase shifts~\cite{Balatsky_2006}. However, in general the phase shift can be treated as a model-dependent parameter. 

In a basic model, where the effects of momentum and spin rotation during scattering event can be thought to be independent of each other, the combined form of the coherence amplitude after scattering is
\begin{eqnarray}
	\Gamma(\mathbf{p}_{\mathrm{out}})i\sigma_y = |S| \mathcal{M} \left[ \gamma(\mathbf{p}_{\mathrm{in}}) i\sigma_y \right] \tilde{\mathcal{M}} &=& |S| e^{-i (\hat{\mathbf{m}} \cdot \bm{\sigma}) \vartheta/2} (\gamma_0 + \bm{\gamma}_{\mathrm{t}}(\mathbf{p}_{\mathrm{in}}) \cdot \bm{\sigma}) e^{-i (\hat{\mathbf{m}} \cdot \bm{\sigma}) \vartheta/2} (i\sigma_y) \nonumber\\ 
&=&	\left[ |S| e^{-i\sigma_z \vartheta} (\gamma_0 + \gamma_z \sigma_z)  + |S|( \gamma_x \sigma_x + \gamma_y \sigma_y) \right] (i\sigma_y),
\end{eqnarray}
where in the last step it has been taken into account that the direction of magnetisation in the solid layer is along the external field, $\hat{\mathbf{m}} \parallel \hat{\mathbf{z}}$. One notes that, first, due to decoupled spin and momentum spaces only the coherent part of momentum reflection from Supplementary Eq.~(\ref{eq:momentum_scatt}) contributes, since the diffuse part of scattering averages the order parameter to zero and magnetically-induced phase shifts are not affecting this average. Second, the $\gamma_x$ and $\gamma_y$ components of triplet $\bm{\gamma}_{\mathrm{t}}$ are not affected by magnetic scattering, meaning that $\displaystyle \vert \uparrow\uparrow \rangle$ and $\displaystyle \vert \downarrow\downarrow \rangle$ pairs ($S_{\mathbf{m}} = \pm 1$) are not magnetically suppressed, since the spins of these pairs scatter with the same phase. Thus, in geometry such as ours, magnetic scattering of this type does not suppress $T_{\mathrm{c}}$ of $^3$He-A further from the value set by momentum scattering. 

For magnetic scattering to have effect, the pairs must be in $S_{\mathbf{m}} = 0$ state, where spins of the $\displaystyle \vert \uparrow \downarrow + \downarrow \uparrow \rangle$ pairs would scatter with different phases. The total phase accumulated by the quasiparticle during the scattering, and thus the energy of surface-bound state, is a combination of rotation of $\gamma_z$ component in spin space and the difference in $\Delta (\mathbf{k})$ due to scattering in momentum space. These two can either enhance or cancel each other's effect on pair breaking. Assuming no pair breaking due to momentum scattering, $S_{\mathbf{m}} = 0$ pairs would be completely broken when the $\gamma_z$ component is rotated by $\vartheta = \pi$ generating zero-energy surface-bound states, which would be the case of resonant magnetic scattering.

The mixing of all pairs ($S_{\mathbf{m}} = 0, \pm 1$) in our experimental configuration becomes possible if the direction of magnetization in the solid layer is random. The average over $\hat{\mathbf{m}}$-angles gives partial suppression of all amplitudes at the surface:
\begin{eqnarray}
	\Gamma(\mathbf{p}_{\mathrm{out}}) i\sigma_y &=& |S| \int \frac{d\Omega_{\hat{\mathbf{m}}}}{4\pi} \mathcal{M}(\hat{\mathbf{m}}) \left[ \gamma(\mathbf{p}_{\mathrm{in}})i\sigma_y \right] \tilde{\mathcal{M}}(\hat{\mathbf{m}}) \nonumber\\ 
&=& \left[ |S| \gamma_0 \cos \vartheta + |S| (\bm{\gamma}_{\mathrm{t}}(\mathbf{p}_{\mathrm{in}}) \cdot \bm{\sigma}) \left(1 - \frac23 \sin^2\frac{\vartheta}{2} \right) \right] (i\sigma_y).
\end{eqnarray}
In this case the magnetic suppression cannot take the total suppression beyond diffuse due to non-negative coefficient for the triplet components.

The only scenario increasing pair breaking enough in this model in order to get more than diffuse suppression of $T_{\mathrm{c}}$ in $^3$He-A, is to allow for correlated spin-orbital scattering, where backscattering quasiparticles experience phase difference due to orbital part of the order parameter alone, and the forward-scattering quasiparticles are affected by magnetic depairing. The latter part would also require a certain orientation of magnetization in the solid layer, inconsistent with the experimental geometry. From the physical angle, this scenario also appears to be unlikely due to absence of plausible spin-orbital coupling mechanism in the layer.  

Another way to formulate the boundary condition for the coherence amplitudes is to model the magnetic layer by a net of polarized or unpolarized scattering centres with potential and exchange interaction $u_0 + J \mathbf{S}_{\mathrm{imp}} \cdot \bm{\sigma}$ with a classical magnetic moment $\mathbf{S}_{\mathrm{imp}}$ of the centres (``impurities'') and liquid-solid exchange coupling coefficient $J$. These act to randomize the directions of scattered quasiparticles, but introduce in general different relaxation times for quasiparticles of different spins. In this modification of Ovchinnikov-Kopnin model of thin dirty layer~\cite{Ovchinnikov_1969,Kopnin_1986} with thickness $d$, one again encounters vanishing diffuse average over directions $\langle \gamma \rangle_\parallel = 0$ that does not further contribute to the magnetic pair breaking. The coherent scattering part is given by spin-dependent relaxation lengths $\ell$, e.g., $\Gamma(\mathbf{p}_{\mathrm{out}}) = \gamma(\mathbf{p}_{\mathrm{in}}) e^{-2d/\ell}$ with ${v^{}_{\mathrm{F}}}/{\ell_\pm} = 2 \pi n^{}_{\mathrm{imp}} N_{\mathrm{F}} (u_0 \pm JS_{\mathrm{imp}})^2$ for $S_z = \pm 1$ pairs, or ${v^{}_{\mathrm{F}}}/{\ell_0} = 2 \pi n^{}_{\mathrm{imp}} N_{\mathrm{F}} (u_0^2 + (JS_{\mathrm{imp}})^2)$ for $S_z = 0$ pairs, where $n^{}_{\mathrm{imp}}$ is the density of scattering centres in the magnetic layer. These expressions are consistent with previous work on bulk systems in aerogel, and they may lead to effects such as $A_1$-$A_2$ phase splitting~\cite{A1-A2_splitting}, but will not result in an excess of zero-energy states because $e^{-2d/\ell}$ coefficients are all positive. The maximal allowed pair breaking is diffuse, $|S|=0$, in the limit $d \gg \ell$ with no magnetic effects discernible.

\subsection{Quantum spin scattering}
\label{sec:spin_exchange}

\begin{figure}[t]
	\includegraphics[width=0.7\textwidth]{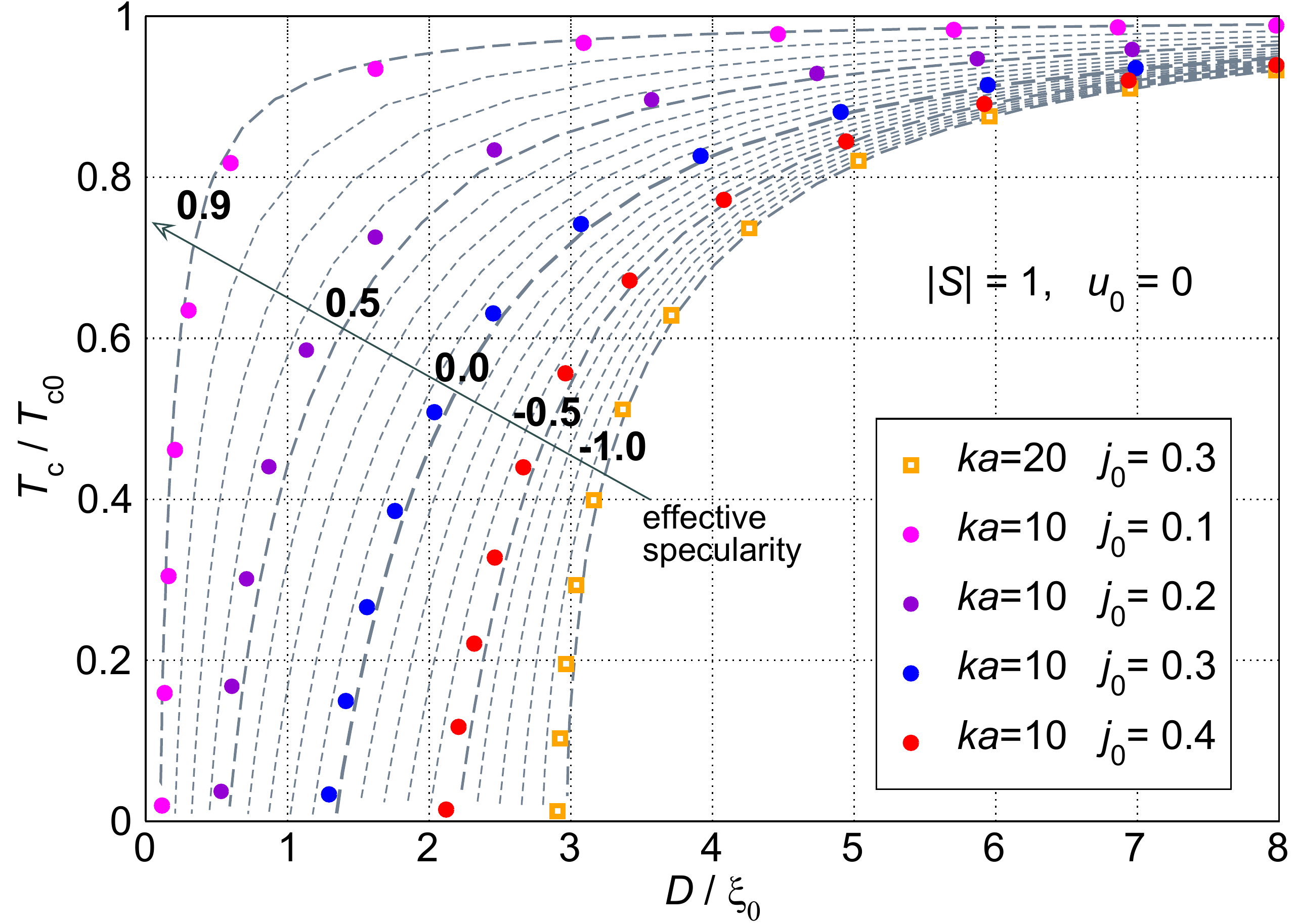}
	\caption{\label{fig:spinTc} \textbf{Suppression of $T_{\mathrm{c}}$ by magnetic scattering on quantum spins.} The momentum scattering is taken to be fully specular, reducing the suppression to be purely from magnetic origin. Dashed lines show the calculated suppression of $T_{\mathrm{c}}$ corresponding to different effective specularities $-1 \leq S_{\mathrm{eff}} \leq 1$. Magnetic scattering, Supplementary Eq.~(\ref{eq:magnetic_scatt}), is parametrized by four scattering phases, $\alpha_{s,t}, \tilde{\alpha}_{s,t}$, in singlet and triplet channels for both particles and holes. The values for these depend on microscopic model of the magnetic layer. Symbols show the resulting suppression corresponding to a set of chosen layer parameters, assuming the orders of magnitude $k \sim k^{}_{\mathrm{F}} \sim 10\,\textrm{nm}^{-1}$ and $a \sim 1\,$nm and no scalar potential in the layer, $u_0 = 0$. We also neglect the dependence of the scattering phases on momentum perpendicular to the wall. The resulting curves can model the suppression over the full range of effective specularity.}
\end{figure}

We propose to extend the boundary conditions to take into account the quantum interference of spin states of the (identical) solid and liquid $^3$He atoms. They will include spin-flip processes and more subtle scattering effects that are missing in the classical magnetic moment approaches. A rough model for the scattering of a liquid $^3$He quasiparticle in the solid layer of thickness $a$ is given by the Hamiltonian in the solid layer
\begin{equation}
	\mathcal{H} = - \frac{\hbar^2}{2m} \bm{\nabla}^2 + U_0 + J \hat{\mathbf{S}}^s \cdot \frac{\hbar}{2}\bm{\sigma},
\end{equation}
where $U_0$ is the potential height of the barrier, $\hat{\mathbf{S}}^s$ is the $\hbar/2$ spin operator of solid layer atom, and $\frac{\hbar}{2}\bm{\sigma}$ is the spin operator of liquid quasiparticle. Below the layer, we set up an impenetrable wall so that all particles are reflected back. The total spin of the quasiparticle and the solid atom participating in scattering can be in the singlet or triplet state, $S_{\mathrm{tot}} = 0,\,1$, with the standard eigenvalues for the product $\langle \hat{\mathbf{S}}^s \cdot \frac{\hbar}{2}\bm{\sigma} \rangle = -\frac34 \hbar^2,\, +\frac14 \hbar^2$. The scattering matrix $e^{-i k z} \to - e^{i\alpha} e^{i k z}$ is (omitting overall $-1$ factor)
\begin{equation}
	\mathcal{M} = 
	\begin{bmatrix} 
		e^{i\alpha_t} & 			& 		    & 		\\
		& e^{i\alpha_s} & 		    & 		\\
		& 			& e^{i\alpha_t} & 		\\
		& 			& 		    & e^{i\alpha_t} 
	\end{bmatrix} 
	\quad 
	\textrm{in total spin basis}
	\quad 
	\begin{pmatrix}
		S_{\mathrm{tot}}=1,& m= +1 \\
		S_{\mathrm{tot}}=0,& m= \; 0  \\
		S_{\mathrm{tot}}=1,& m= \; 0 \\
		S_{\mathrm{tot}}=1,& m= -1 \\
	\end{pmatrix}
	,
\end{equation}
where the phases of the singlet and triplet scattering channels are given by 
\begin{equation}
	\label{eq:part_phase}
	\tan \frac{\alpha_s}{2} = \frac{\tan\left( ka \sqrt{1 - u_0 + \frac34 j_0} \right)}{\sqrt{1 - u_0 + \frac34 j_0}}
	\;,\qquad
	\tan \frac{\alpha_t}{2} = \frac{\tan\left( ka \sqrt{1 - u_0 - \frac14 j_0} \right)}{\sqrt{1 - u_0 - \frac14 j_0}}
\end{equation}
with definitions  
\begin{equation}
	\label{eq:param}
	u_0 = \frac {2m U_0}{\hbar^2 k^2}
	\;,\qquad
	j_0 = \frac {2m J}{k^2} \,.
\end{equation}

We transform to the basis labelled by $z$-projections of liquid ($\sigma,\sigma'$) and solid ($S^s_z, S_z^{s'}$) spins before and after the scattering event using Clebsch-Gordan matrix $\mathcal{C}$,
\begin{equation}
	\vert \sigma, S^s_z \rangle = 
	\begin{pmatrix} 
		+1/2, +\frac12 \\
		+1/2, -\frac12 \\
		-1/2, +\frac12 \\
		-1/2, -\frac12 
	\end{pmatrix} 
	= {\mathcal{C}}^{-1}
	\begin{pmatrix}
		S_{\mathrm{tot}}=1,& m= +1 \\
		S_{\mathrm{tot}}=0,& m= \;0  \\
		S_{\mathrm{tot}}=1,& m= \;0 \\
		S_{\mathrm{tot}}=1,& m= -1 \\
	\end{pmatrix} 
	, \quad
	\mathcal{C} = 
	\begin{bmatrix} 
		1		  & 			& 		    & 		\\
		& \frac{1}{\sqrt{2}}    & -\frac{1}{\sqrt{2}}    & 		\\
		& \frac{1}{\sqrt{2}} 	&  \frac{1}{\sqrt{2}}    & 		\\
		& 			& 		    & 1 
	\end{bmatrix} 
	,
\end{equation}
to get the scattering matrix 
\begin{eqnarray}
	\mathcal{M}_{\sigma,\sigma' \otimes S^s_z,S_z^{s'}} &=& {\mathcal{C}}^{-1}
	\begin{bmatrix} 
		e^{i\alpha_t} & 			& 		    & 		\\
		& e^{i\alpha_s} & 		    & 		\\
		& 			& e^{i\alpha_t} & 		\\
		& 			& 		    & e^{i\alpha_t} 
	\end{bmatrix} 
	\mathcal{C} \nonumber\\
&=& e^{i \alpha_+/2}
	\left( \begin{array}{cc|cc}
		e^{- i \alpha_-/2} & 				  &  					&			 \\   
		& \cos\frac{\alpha_-}{2} & -i\sin\frac{\alpha_-}{2} 	&			 \\ 
		\hline
		& -i\sin\frac{\alpha_-}{2} & \cos\frac{\alpha_-}{2}  &  			 \\
		& 				&					& e^{- i \alpha_-/2} 
	\end{array} \right)
	\raisebox{-6mm}{$\Big|_{\alpha_\pm = \alpha_s \pm \alpha_t}$ }
	\,.
\end{eqnarray}
The separate $2\times 2$ blocks correspond to ($\sigma,\sigma'$) space; within each block we have $2\times 2$ ($S^s_z,S_z^{s'}$) space. One gets similar expression for the hole sector: 
\begin{equation}
	\tilde{\mathcal{M}}_{\sigma,\sigma' \otimes S^s_z,S_z^{s'}} = (-i \sigma_y) {\mathcal{C}}^{-1}
	\begin{bmatrix} 
		e^{i\tilde \alpha_t} & 			& 		    & 		\\
		& e^{i\tilde \alpha_s} & 		    & 		\\
		& 			& e^{i\tilde \alpha_t} & 		\\
		& 			& 		    & e^{i\tilde \alpha_t} 
	\end{bmatrix} 
	\mathcal{C} (i \sigma_y) \,.
\end{equation}
Here $(i \sigma_y)$-factors act on liquid spin variables, and the scattering phases 
\begin{equation}
	\label{eq:hole_phase}
	\tan \frac{\tilde\alpha_s}{2} = -\frac{\tan\left( ka \sqrt{1 - u_0 - \frac34 j_0} \right)}{\sqrt{1 - u_0 - \frac34 j_0}}
	\;,\qquad
	\tan \frac{\tilde\alpha_t}{2} = -\frac{\tan\left( ka \sqrt{1 - u_0 + \frac14 j_0} \right)}{\sqrt{1 - u_0 + \frac14 j_0}}
\end{equation}
do not show the particle-hole symmetry, which was present in the scattering on a classical spin, due to difference between singlet and triplet factors $-3/4,+1/4$. (In large-spin or particle-hole-symmetric case, we would have $\tilde{\alpha}_s = - \alpha_t, \tilde{\alpha}_t = - \alpha_s$.)

By taking average of solid spin configurations, we obtain the boundary condition for the coherence amplitudes: 
\begin{equation}
	\Gamma_{\alpha\beta}(\mathbf{p}_{\mathrm{out}}) = \sum_{S^s_z=\pm \frac12} \mathcal{P}(S^s_z) \, \langle S^s_z \vert \mathcal{M}_{\alpha, \sigma\otimes S^s_z,S_z^{s'}} \; \gamma^{}_{\sigma \sigma'}(\mathbf{p}_{\mathrm{in}}) \; \tilde{\mathcal{M}}_{\sigma', \beta \otimes S_z^{s'},S^s_z} \vert S^s_z \rangle \,.
\end{equation}
In the assumption of unpolarized solid layer, $\mathcal{P}(+\frac12) = \mathcal{P}(-\frac12) = 1/2$ (i.e., the NMR field is not strong enough to orient significant fraction of the spins), the triplet components upon scattering from the magnetic layer are multiplied by an additional factor:
\begin{equation}
	\label{eq:magnetic_scatt}
	\begin{split}
	\Gamma_{x,y,z} (\mathbf{p}_{\mathrm{out}}) &= Q_{\mathrm{sf}} \; |S| \;  \gamma_{x,y,z}(\mathbf{p}_{\mathrm{in}})
	\\
	Q_{\mathrm{sf}} &= e^{i (\alpha_+ + \tilde\alpha_+) / 2 }\; \frac12 \left( e^{-i \alpha_-/2} \cos \frac{\tilde\alpha_-}{2} + e^{-i \tilde\alpha_-/2} \cos \frac{\alpha_-}{2}\right) 
	\\
	& = \frac14 \left( e^{i(\alpha_s+ \tilde\alpha_t)} + e^{i(\tilde\alpha_s + \alpha_t)} + 2 e^{i(\alpha_t+ \tilde\alpha_t)} \right) 
	\,.
	\end{split}
\end{equation}
Complex magnetic pair-breaking parameter $Q_{\mathrm{sf}}$ is a mixture of singlet and triplet channel scattering phases of both particles and holes. It can easily become negative, depending on the specific values of the phases in Supplementary Eqs.~(\ref{eq:part_phase}) and~(\ref{eq:hole_phase}) which in this model are functions of solid layer parameters $ka$ and $j_0$ defined in Supplementary Eq.~(\ref{eq:param}). We assume partially specular momentum scattering with less than diffuse $T_{\mathrm{c}}$ suppression. In this situation the largest increase in the density of zero-energy surface-bound states arises when $\alpha_s + \tilde{\alpha}_t \sim \pi$, $\tilde{\alpha}_s + \alpha_t \sim \pi$, and $\alpha_t + \tilde{\alpha}_t \sim \pi$, so that the real part $\Re Q_{\mathrm{sf}} = \left( \cos (\alpha_s+ \tilde{\alpha}_t) + \cos(\tilde{\alpha}_s + \alpha_t) + 2 \cos(\alpha_t + \tilde{\alpha}_t) \right) /4 \sim -1$. Further adjustment of the parameters can lead to total $T_{\mathrm{c}}$ suppression corresponding to any value of effective specularity $-|S| \leq S_{\mathrm{eff}} \leq |S|$. Thus, the suppression of $T_{\mathrm{c}}$ can be more than diffuse even for fully specular momentum scattering $|S| = 1$, in which case the magnetic pair breaking alone covers the full range of suppressions matching the effective specularity from specular to retroreflective (Supplementary Fig.~\ref{fig:spinTc}).

In order to get $T_{\mathrm{c}}$ suppression matching effective specularity $S_{\mathrm{eff}} = -0.4$, as detected with solid $^3$He surface boundary layer (see Fig.~3a in the main text), the underlying specularity of momentum scattering needs to be $S \ge 0.4$. This can in principle be possible if the atomically smooth silicon surfaces of the cavity~\cite{Zhelev_RevSciInst2018} are not significantly roughened by solid $^3$He layer, and if atomic-level roughness of the solid substrate itself can promote specular scattering.

The three mechanisms capable of producing $T_{\mathrm{c}}$ suppression more than diffuse, as discussed here, are all illustrated in Fig.~3b in the main text.

\clearpage

\section{Temperature correction}
\label{sec:T_correction}

We measure the temperature $T_{\mathrm{Ag}}$ of the silver plate on which the sample container sits. Due to thermal boundary resistance $R_{\mathrm{K}}$ across the sintered silver heat exchanger, there is a temperature gradient $\Delta T$ between the sample helium and the silver plate: $\Delta T = T_{\mathrm{He}} - T_{\mathrm{Ag}} = R_{\mathrm{K}} \dot{Q}$, where $T_{\mathrm{He}}$ is the helium temperature and $\dot{Q} = \dot{Q}_0 + \dot{Q}_{\mathrm{NMR}}$ is the heat flux into the sample~\cite{Kapitza_res}. Here residual constant heat flux is $\dot{Q}_0$ and the heat flux due to NMR pulses is $\dot{Q}_{\mathrm{NMR}}$. Typical form of boundary resistance is $R_{\mathrm{K}} = \frac{K}{T^\alpha}$, where $K$ is a constant dependent on material and boundary properties~\cite{Pobell}. At high temperatures we expect $\alpha = 3$, following from the acoustic mismatch of phonons at the interface. However, at low temperatures below 10\,mK, $\alpha$ typically decreases towards unity and is different for pure $^3$He and for mixtures of $^4$He and $^3$He~\cite{Kapitza_lowT}. Therefore, we take $\alpha$ as a free parameter for each surface preplating investigated. 

\begin{figure*}
	\includegraphics[width=\textwidth]{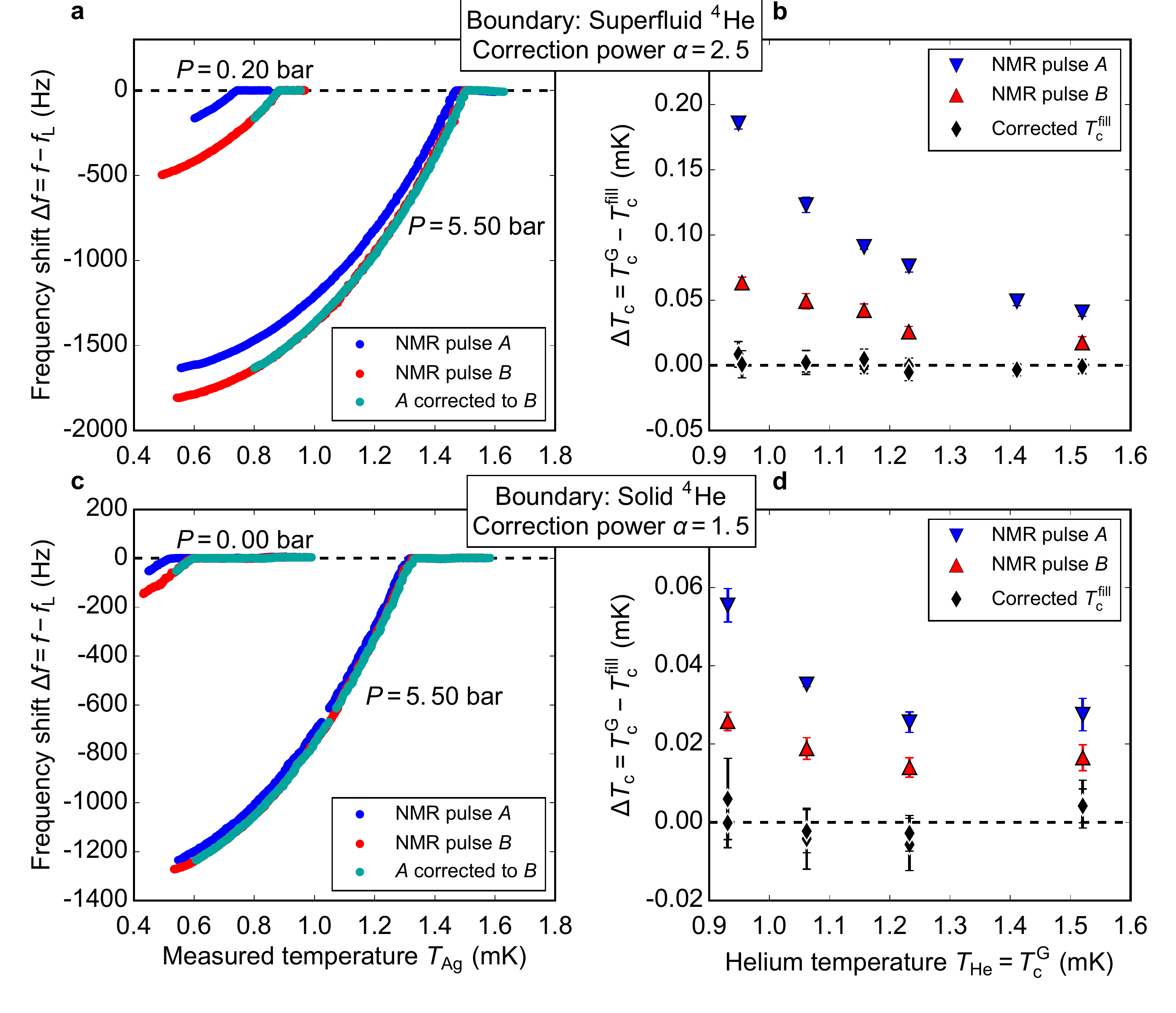}
	\caption{\label{fig:NMR_calib} \textbf{Calibration of heating caused by NMR pulses.} \textbf{a, c,} Frequency shifts in the cavity at two pressures with ``specular'' and ``diffuse'' boundary conditions for NMR pulses $A$ and $B$. Correction between pulses has been done using Supplementary Eq.~(\ref{eq:temp_corr_diff}). \textbf{b, d,} Difference between measured superfluid transition temperatures $T_{\mathrm{c}}^{\mathrm{fill}}$ in the fill line bulk marker and the literature value $T_{\mathrm{c}}^{\mathrm{G}}$ at different pressures corresponding to values of $T_{\mathrm{c}}^{\mathrm{G}}$ between 0.9 and 1.6\,mK. Each upwards and downwards pointing triangle represents the average of several measured bulk transitions. Corrected temperatures are reached using Supplementary Eq.~(\ref{eq:temp_corr_final}).}
\end{figure*}

We fit $\alpha$ by measuring the change in $\Delta T$ as a function of $\dot{Q}$ across the heat exchanger over the whole temperature range of interest. This is done by performing full temperature sweeps using two different-sized NMR pulses denoted as pulse $A$ and pulse $B$, where the total rf-field power generated by pulse $A$ is ten times the power generated by pulse $B$. These pulses correspond to tipping angles $\beta \approx 10^\circ$ and $\beta \approx 3^\circ$, respectively. We denote the two significantly different NMR-induced heat fluxes into the sample by $\dot{Q}_{\mathrm{NMR},A}$ and $\dot{Q}_{\mathrm{NMR},B}$, which result in two different dependences of the superfluid frequency shift in the cavity on the measured silver-plate temperature (Supplementary Fig.~\ref{fig:NMR_calib}a,c). Since any measured frequency shift $\Delta f$ in the superfluid $^3$He-A in the cavity, using small-tipping-angle pulses, always corresponds to the same helium temperature, we can calibrate the additional heating using equation
\begin{equation}
\label{eq:temp_corr}
\int_{T_{\mathrm{Ag}}}^{T_{\mathrm{He}}} \frac{dT}{R_{\mathrm{K}}} = \frac{T_{\mathrm{He}}^{\alpha + 1} - T_{\mathrm{Ag}}^{\alpha + 1}}{(\alpha + 1) K} = \dot{Q}.
\end{equation}
From this it follows for any $\Delta f$
\begin{equation}
\label{eq:temp_corr_diff}
\dot{Q}_A - \dot{Q}_B = \dot{Q}_{\mathrm{NMR},A} - \dot{Q}_{\mathrm{NMR},B} = \dot{Q}_{\mathrm{NMR},A-B} = \frac{T_{\mathrm{Ag},B}^{\alpha + 1} - T_{\mathrm{Ag},A}^{\alpha + 1}}{(\alpha + 1) K}.
\end{equation}
Since $K$ and $\dot{Q}_{\mathrm{NMR},A-B}$ are both constants, we can fit the measured temperature differences corresponding to each frequency shift to determine $\alpha$. We find it to be independent of both temperature and pressure over the 0.5--1.5\,mK, 0.0--5.5\,bar range. With ``specular'' boundary condition $\alpha=2.5$ and with ``diffuse'' $\alpha=1.5$. These values collapse data of different pulses, see Supplementary Fig.~\ref{fig:NMR_calib}a,c. We have also compared the frequency shifts using pulse $B$ and a medium-sized pulse having four times the power generated by $B$. The frequency shifts match within the experimental error, leading us to conclude that the heating caused by pulse $B$ is negligible, i.e., $\dot{Q}_{\mathrm{NMR},B} \approx 0$.

Given the determination of $\alpha$ for each boundary condition, we can make the final correction for residual heat leak into the sample, $\dot{Q}_0$. This is done by comparison of the measured silver plate temperature $T_{\mathrm{Ag}} \equiv T_{\mathrm{c}}^{\mathrm{fill}}$ during the superfluid transition in the fill line bulk marker, which is most directly connected to the heat exchanger, with the literature value given by Greywall, $T_{\mathrm{He}} \equiv T_{\mathrm{c}}^{\mathrm{G}}$~\cite{Greywall_1986}. The comparison is made at each pressure for both pulse $A$ and pulse $B$ (Supplementary Fig.~\ref{fig:NMR_calib}b,d). Referring to Supplementary Eq.~(\ref{eq:temp_corr}), this procedure determines
\begin{eqnarray}
\label{eq:CA_CB}
C_A &=& T_{\mathrm{c,G}}^{\alpha + 1} - T_{\mathrm{c,fill},A}^{\alpha + 1} = (\alpha+1) \left( \dot{Q}_0 + \dot{Q}_{\mathrm{NMR},A} \right) K \nonumber\\
C_B &=& T_{\mathrm{c,G}}^{\alpha + 1} - T_{\mathrm{c,fill},B}^{\alpha + 1} = (\alpha+1) \dot{Q}_0 K,
\end{eqnarray}
where we use the fact that the heating due to pulse $B$ is negligible. Then, the conversion of any measured temperature $T_{\mathrm{Ag}}$ to actual helium temperature $T_{\mathrm{He}}$ is
\begin{equation}
\label{eq:temp_corr_final}
T_{\mathrm{He}} = \left( T_{\mathrm{Ag}}^{\alpha + 1} + C_{A} \right)^{1/ \left( \alpha + 1 \right)},
\end{equation}
with a similar expression for pulse $B$.

\begin{figure}
	\includegraphics[width=\textwidth]{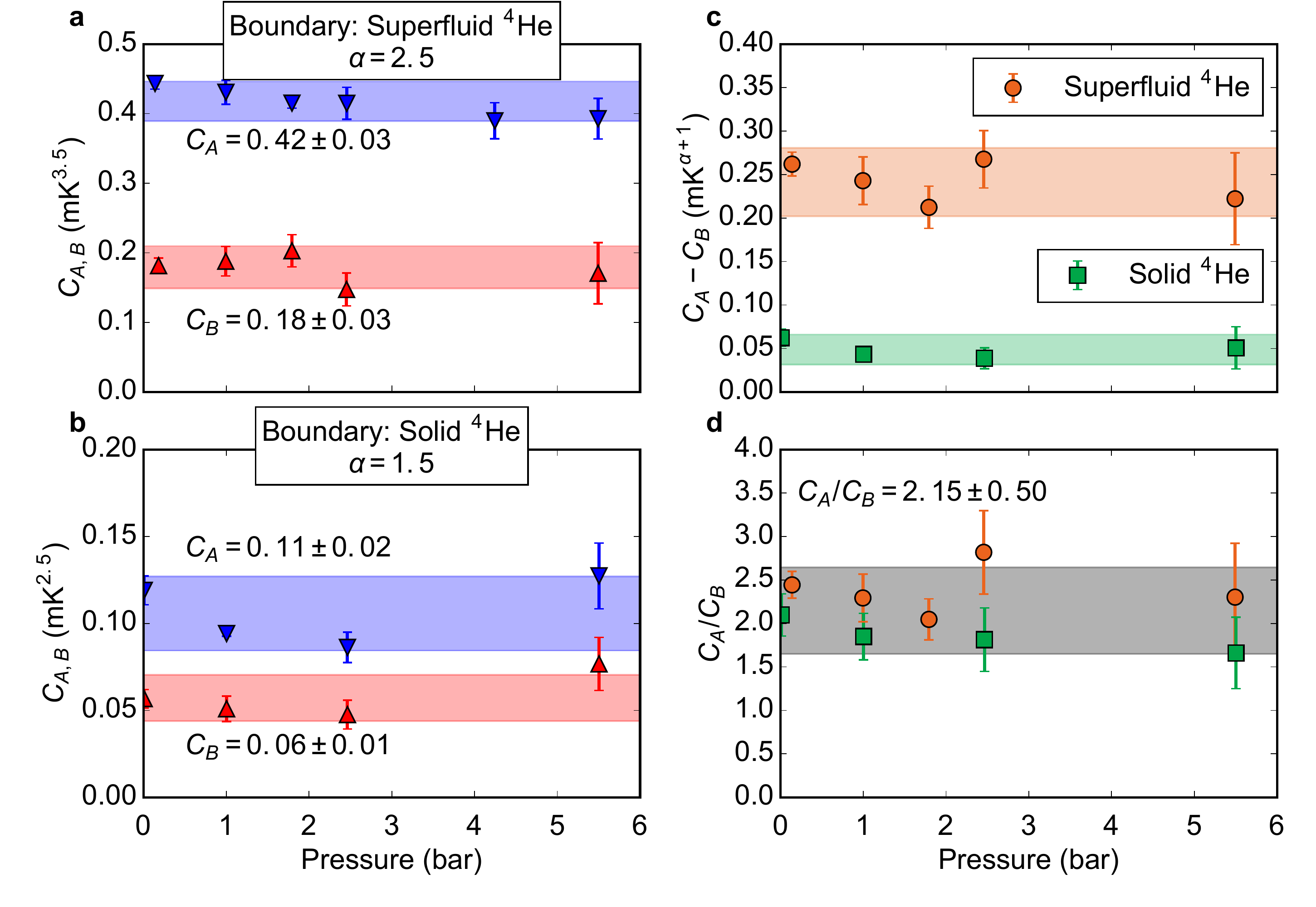}
	\caption{\label{fig:NMR_corr} \textbf{Model parameters used in temperature correction.} \textbf{a, b,} Constants $C_A$ and $C_B$ (Supplementary Eq.~(\ref{eq:CA_CB})) with two boundary conditions as defined at bulk superfluid transitions in the fill line. Each point is an average of several measured $T_{\mathrm{c}}^{\mathrm{fill}}$, same as in Supplementary Fig.~\ref{fig:NMR_calib}b,d. The values of $C_A$ and $C_B$ are independent of pressure and significantly smaller with ``diffuse'' boundary condition due to smaller ``poisoning'' of heat exchanger by the thinner $^4$He film. \textbf{c,} Difference between constants $C_A$ and $C_B$ as a function of pressure with both boundary conditions, giving the correction between the pulses. \textbf{d,} The ratio $C_A/C_B = 1 + \dot{Q}_{\mathrm{NMR},A}/\dot{Q}_0$ is defined from the same measurements of $T_{\mathrm{c}}^{\mathrm{fill}}$ and does not depend on the boundary condition or the pressure, as expected. The coloured bands in each panel indicate the values averaged over pressure with the widths of the bands showing the uncertainties.}
\end{figure}

The results for $C_A$ and $C_B$ (Supplementary Fig.~\ref{fig:NMR_corr}) show pressure independence. Therefore, the values are averaged over pressure in order to correct all temperatures in both the main text and the Supplementary Information. The corrected values of bulk transition temperatures in fill line are shown in Supplementary Fig.~\ref{fig:NMR_calib}b,d. The usage of constant $\alpha$ and universal pressure-independent $C_A$ or $C_B$ satisfactorily corrects all the measurements within error limits from $T_{\mathrm{c}}^{\mathrm{G}}$. The uncertainty limits of $C_A$ and $C_B$ increase the uncertainty of the corrected temperatures as compared to the small error in the original measured temperatures.

We assume that the upper bound of the boundary resistance with a solid $^3$He surface boundary layer is set by that of a solid $^4$He layer in order to quantify uncertainties in temperature in this case.

\clearpage

\bibliography{200nm_slab}
\bibliographystyle{naturemag}

\end{document}